\documentclass[fleqn,usenatbib]{mnras}

\usepackage{newtxtext,newtxmath}

\usepackage[T1]{fontenc}
\usepackage{ae,aecompl}

\usepackage{graphicx}	
\usepackage{amsmath}	
\usepackage{amssymb}	

\usepackage{enumerate}
\usepackage{color}
\usepackage{graphicx}
\usepackage{multicol}
\usepackage{savesym}
\usepackage{color}
\usepackage{tipa}
\usepackage{csvsimple}
\usepackage{arydshln}
\usepackage{arydshln}

\newcommand{\perbeam}{$\,{\rm beam}^{-1}$}

\newcommand{\micro}{\text{$\mu$}}
\newcommand{\milli}{\text{m}}
\newcommand{\centi}{\text{c}}
\newcommand{\kilo}{\text{k}}

\newcommand{\hertz}{\text{Hz}}
\newcommand{\GHz}{\text{GHz}}

\newcommand{\yr}{\text{yr}}
\newcommand{\watt}{\text{W}}
\newcommand{\jansky}{\text{Jy}}
\newcommand{\ujy}{\micro\text{Jy}}
\newcommand{\meter}{\text{m}}
\newcommand{\parsec}{\text{pc}}
\newcommand{\second}{\text{s}}
\newcommand{\erg}{\text{erg}}

\newcommand{\kev}{\text{keV}}

\newcommand{\guass}{\text{G}}
\newcommand{\ergs}{\text{erg s$^{-1}$}}

\newcommand{\Kelvin}{\text{K}}
\newcommand{\hour}{\text{hr}}
\newcommand{\E}[1]{\times10^{#1}}
\newcommand{\Msol}{{\rm M}_{\odot}}
\newcommand{\Rsol}{{\rm R}_{\odot}}
\newcommand{\U}[3]{$#1_{-#3}^{+#2}$}

\newcommand{\flux}{\text{erg cm$^{-2}$ s$^{-1}$}}

\newcommand{\barcs}{\ensuremath{.\hspace{-1.1mm}^{\prime\prime}}}
\newcommand{\bmins}{\ensuremath{.\hspace{-1.2mm}^{\prime}}}
\newcommand{\bsecs}{\ensuremath{.\hspace{-1.2mm}^{s}}}
\newcommand{\bdegs}{\ensuremath{.\hspace{-1.2mm}^{\circ}}}

\title[NGC\,300 microquasar]
  {A newly discovered double-double candidate microquasar in NGC\,300}

\author[R. Urquhart et al.]
{R.~Urquhart,$^{1,2}$\thanks{Emails: ryan.urquhart@icrar.org (RU)}
R.~Soria,$^{3,1,4}$
M.W.~Pakull,$^{5}$
J.C.A.~Miller-Jones,$^{1}$
G.E.~Anderson,$^{1}$\newauthor
R.M.~Plotkin,$^{1}$
C.~Motch,$^{5}$
T.J.~Maccarone,$^{6}$
A.F.~McLeod$^{7}$
and S.~Scaringi$^{7}$
\\
$^{1}$International Centre for Radio Astronomy Research, Curtin University, GPO Box U1987, Perth, WA 6845, Australia\\
$^{2}$Center for Data Intensive and Time Domain Astronomy, Department of Physics and Astronomy, Michigan State University,\\\qquad East Lansing, MI, USA\\
$^{3}$College of Astronomy and Space Sciences, University of the Chinese Academy of Sciences, Beijing 100049, China\\
$^{4}$Sydney Institute for Astronomy, School of Physics A28, The University of Sydney, Sydney, NSW 2006, Australia\\
$^{5}$Observatoire astronomique, Universit\'e de Strasbourg, CNRS, UMR 7550, 11 rue de l'Universit\'e, 67000, Strasbourg, France\\
$^{6}$Physics Department, Texas Tech University, PO Box 41051, Lubbock, TX 79409, USA\\
$^{7}$School of Physical and Chemical Sciences, University of Canterbury, Christchurch, New Zealand\\
\\
}

\date{Accepted XXX. Received YYY; in original form ZZZ}

\pubyear{2018}

\begin{document}
\label{firstpage}
\pagerange{\pageref{firstpage}--\pageref{lastpage}}
\maketitle

\begin{abstract}
We present the discovery of a powerful candidate microquasar in NGC\,300, associated with the S\,10 optical nebula (previously classified as a supernova remnant). {\it Chandra} images show four discrete X-ray knots aligned in the plane of the sky over a length of $\approx$150 pc. The X-ray emission from the knots is well fitted with a thermal plasma model at a temperature of $\approx$0.6 keV and a combined 0.3--8 keV luminosity of $\approx$10$^{37}$ erg s$^{-1}$. The X-ray core, if present at all, does not stand out above the thermal emission of the knots: this suggests that the accreting compact object is either currently in a dim state or occulted from our view. We interpret the emission from the knots as the result of shocks from the interaction of a jet with the interstellar medium (possibly over different epochs of enhanced activity). Cooler shock-heated gas is likely the origin also of the optical bubble and lobes near the X-ray structure, detected in images from the {\it Hubble Space Telescope} and the Very Large Telescope. In the radio bands, we observed the region with the Australia Telescope Compact Array, and discovered an elongated radio nebula (about 170 $\times$ 55 pc in size) with its major axis aligned with the chain of {\it Chandra} sources. The radio nebula has an integrated $5.5$ GHz radio luminosity of $\approx$10$^{34}\,\ergs$ for a distance of 1.88 Mpc.
The morphology, size and luminosity of the extended X-ray, optical and radio structure suggest that NGC\,300-S\,10 belongs to the same class of powerful ($P_{\rm jet} > 10^{39}$ erg s$^{-1}$) microquasars as SS\,433, Ho II X-1 and NGC\,7793-S\,26.
\end{abstract}


\section{Introduction}

Collimated jets can carry away as kinetic energy a significant fraction of the total accretion power of a compact object \citep[e.g.,\,][]{1999PhR...311..225L,2004MNRAS.355.1105F,2006MNRAS.372...21A,2013ApJ...762..103K,2014Natur.515..376G}. This energy is imparted into the surrounding gas, and often creates large-scale structures such as bubbles, lobes and hotspots, detected both around supermassive black holes \citep[e.g.,\,][]{1984ApJ...285L..35P,1984RvMP...56..255B,1998MNRAS.296..445H,2005ApJ...622..797K} and around off-nuclear (stellar-mass) compact objects
\citep[e.g.,\,][]{1992Natur.358..215M,1998AJ....116.1842D,1999ARA&A..37..409M,2002Sci...298..196C,2005Natur.436..819G, 2010Natur.466..209P}. In the case of supermassive systems, the injection of mechanical power has substantial feedback effects on the evolution of the host galaxy and the growth of the black hole itself \citep[e.g.,\,][]{2012ARA&A..50..455F,2012NJPh...14e5023M,2013Sci...341.1082M,2015ARA&A..53..115K}. In stellar-mass systems (often referred to as ``microquasars''), jets may not be powerful enough to affect the global structure of their host galaxy; however, they still impart significant kinetic energy into the interstellar medium (ISM) and their cumulative effect can influence the evolution of galactic magnetic fields \citep{2005MNRAS.360.1085F, 2008ApJ...686.1145H}. Thus, the most powerful microquasars in the Milky Way and nearby galaxies are excellent laboratories for studies of how jets impact their surroundings. An advantage of microquasars is that, among them, we find the nearest examples of steady, highly super-Eddington accretion in the local universe \citep{2004ASPRv..12....1F,2007MNRAS.377.1187P,2014Sci...343.1330S}. Instead, most nuclear supermassive black holes grow at or below the Eddington limit \citep[e.g.,\,][]{2010MNRAS.402.2637S,2010MNRAS.408L..95K,2012MNRAS.425..623L} except perhaps during the heavily obscured early phase of quasar growth \citep[e.g.,\,][]{2003ApJ...596L..27K,2014ApJ...784L..38M,2015ApJ...804..148V,2017ApJ...836L...1T}, or during transient episodes of tidal disruption events \citep[e.g.,\,][]{2009MNRAS.400.2070S,2014ApJ...781...82C,2016ApJ...819L..25A}.

There is both observational and theoretical evidence for jets from candidate super-Eddington compact accretors in ultraluminous X-ray sources (ULXs: \citealt{2011NewAR..55..166F,2017ARA&A..55..303K}). Magneto-hydrodynamic (MHD) simulations \citep[e.g.,\,][]{2005ApJ...628..368O,2011ApJ...736....2O,2014ApJ...796..106J,2015MNRAS.453.3213S,2017PASJ...69...92K,2017MNRAS.469.2997N} indicate that powerful outflows are launched from the inner part of the super-critical disk, where radiation pressure forces dominate over gravitational forces; the outflows generate a lower-density polar funnel, inside which a relativistic jet is collimated and accelerated. However, it is unclear whether or not this polar funnel and associated jets are a necessary component of every system in the super-critical regime (regardless of the mass, spin, magnetic field and nature of the compact object) or instead there are alternative solutions with and without a relativistic jet. For example, if the super-critical accretor is a strongly magnetized neutron star, the inner disk may be truncated at the magnetospheric radius and the thick outflow funnel may not form \citep{2015MNRAS.454.2539M}. It is also still actively debated \citep{2012MNRAS.419L..69N,2013ApJ...762..104S,2013MNRAS.431..405R,2014MNRAS.439.1740M} whether jets are primarily powered by the spin-down of the compact object via a Blandford-Znajek process \citep{1977MNRAS.179..433B,2011MNRAS.418L..79T,2015MNRAS.454L...6M} or instead by the accretion power released in the disk \citep{1982MNRAS.199..883B}.

One example of a super-Eddington source with collimated jets is Ho II X-1. This jetted ULX ($L_x >10^{40}\,\ergs$) has a distinct triple radio structure: two outer knots from a previous outburst of expanding ejecta and a third inner knot resulting from a more recent ($\lesssim10\,\yr$) ejection event \citep{2014MNRAS.439L...1C,2015MNRAS.452...24C}. The inner central knot also appears variable, fading by a factor of at least 7.3 over $1.5\,\yr$ \citep{2015MNRAS.452...24C}. There must be significant energy being imparted into the ISM to inflate the Ho II X-1 radio nebula \citep{2012ApJ...749...17C}. \citet{2015MNRAS.452...24C} argue that the Ho II X-1 radio nebula is inflated by flaring events from a transient jet, rather than a continuous jet. In several other cases, ULXs are surrounded by large ($\approx$100--300 pc in diameter) bubbles of shock-ionized gas \citep{2002astro.ph..2488P,2008AIPC.1010..303P}, but there is no direct evidence to attribute the mechanical power to a collimated jet rather than a broad outflow.

At slightly lower radiative luminosities, we have the candidate super-Eddington microquasar M\,83-MQ1 \citep{2014Sci...343.1330S}. Unlike Ho II X-1, the X-ray emission from the central source ($L_X\approx7\times10^{37}\,\ergs$) is not sufficient enough to exceed the ULX threshold ($L_X \sim10^{39}\,\ergs$ and above). However, the mechanical power output ($P_{\rm jet} \approx 3\E{40}\,\ergs$), inferred from optical and infrared diagnostic emission lines, places the source in the super-Eddington regime.

Another example of a powerful (candidate super-Eddington) off-nuclear microquasar with even lower core radiative luminosity, is S\,26 in the nearby galaxy NGC\,7793 \citep{2010Natur.466..209P,2010MNRAS.409..541S}. S\,26 consists of a hard non-thermal X-ray core, two radio and X-ray hotspots, radio lobes, and a radio/optical cocoon with a projected size of $\approx$300 $\times$ 150 pc. The hotspots are thought to be termination shocks as the jet impacts the ISM, with the X-ray emission (thermal plasma) and radio emission (synchrotron) coming from different physical processes \citep{2010MNRAS.409..541S}. While the core X-ray luminosity ($L_X \approx 6\E{36}\,\ergs$) is relatively low, \citet{2010Natur.466..209P} find a time-averaged mechanical power output of $P_{\rm jet}\sim$ a few $\times 10^{40}\,\ergs$, suggesting that super-Eddington accretion is indeed taking place, for a stellar-mass accretor ($\lesssim100\,\Msol$). Clearly, for sources like S\,26 and M\,83-MQ1, the impact on their surroundings is longer-lived than their radiative powers.

Both S\,26 and M83-MQ1 are likely more powerful analogues of the Galactic microquasar SS\,433 ($L_X \sim10^{36}\,\ergs$), whose powerful jet ($P_{\rm jet}\gtrsim 10^{39}\,\ergs$; \citealt{1980MNRAS.192..731Z,2004ASPRv..12....1F,2007A&A...463..611B,2011MNRAS.414.2838G,2017MNRAS.467.4777F}) is observed interacting with the surrounding bubble, W50, inflating the radio `ears' \citep{1998AJ....116.1842D} and creating X-ray hotspots \citep{2007A&A...463..611B}. In all three of these sources the mechanical power dominates over the observed radiative power, although it is possible that most of the directly emitted X-ray photons are occulted from our view by optically thick material in the disk plane, if the sources are seen at high inclination. These microquasars, along with the ULX Ho II X-1, are all distinct jet sources and it is currently unclear what properties of the central source and ISM leads to such a diverse range of observables. A larger sample of super-Eddington microquasars is required to understand how these objects impart their mechanical energy into the surrounding medium, what fraction of such sources have collimated jets, how the jet depends on compact object type/mass, and how the jet properties relate to those of jets in sub-Eddington microquasars.

Based on our early optical observational efforts to understand the nature of ultraluminous X-ray sources in nearby galaxies \citep{2002astro.ph..2488P,2003RMxAC..15..197P} we noted that several ULX nebulae had indeed previously been catalogued (e.g.,\,\citealt{1997ApJS..108..261B, 1997ApJS..112...49M, 1997ApJS..113..333M}) as unusually extended optically selected candidates for supernova remnants. Some outstanding examples previously reported or presently studied by the Strasbourg members of the present collaboration include Holmberg IX X-1 \citep{1995ApJ...446L..75M,2011ApJ...734...23G}, M81 X-6 (SNR \#22/23, \citealt{1997ApJS..112...49M, 2003RMxAC..15..197P}), NGC\,5585 X-1 (SNR \#1, \citealt{1997ApJS..112...49M}; Soria et al in prep), NGC\,2403 X-1 (SNR \#14/15, \citealt{1997ApJS..113..333M}; Pakull et al. in prep) and microquasar S26 in NGC\,7793 \citep{1997ApJS..108..261B,2010Natur.466..209P,2010MNRAS.409..541S}. As mentioned earlier, the latter source is not ultraluminous in X-rays, but emits ultraluminous mechanical power. Moreover, the microquasar S\,26 displays a linear triple X-ray source morphology reminiscent of a much larger radio galaxy like Cyg A with its central black hole and (facing) X-ray/radio hot spots. We recall here that also the Galactic jet-source SS\,433 and its radio/X-ray nebula W\,50 would display such a triple point-like source morphology (e.g.,\,\citealt{2011MNRAS.414.2838G}) if observed at a distance of $\sim$ a few Mpc.

\begin{figure}
\hspace{-0.5cm}
\includegraphics[width=0.52\textwidth]{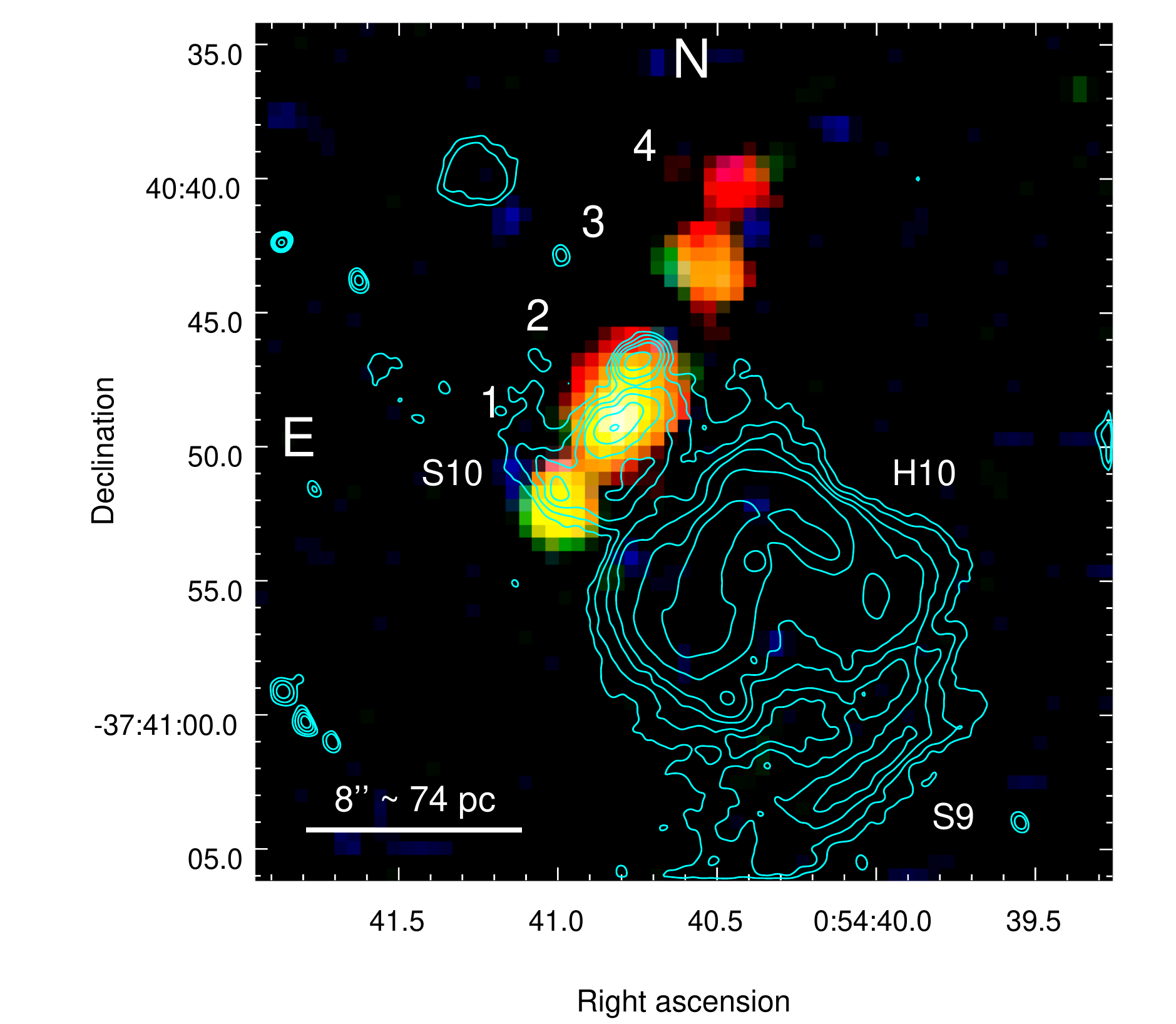}\\
\vspace{-0.4cm}
 \caption{Stacked {\it Chandra} ACIS-I image of NGC\,300 S\,10, showing 4 aligned sources; we refer to those sources as ``knots'' 1--4, throughout this paper, as labelled in this image. Red represents 0.3--1.0 keV, green 1--2 keV, and blue 2--7 keV. The image has been adaptively smoothed with a Gaussian kernel of $\sigma=1$ pixel and has a pixel size of 0\barcs492. Cyan contours show the smoothed VLT continuum-subtracted H$\alpha$ intensity map; contour levels are arbitrary and for visualisation purposes. The complex structure of the H$\alpha$ emission can be divided into the shock-ionized S\,10 region (associated with the X-ray knots), a photo-ionized H II region (H\,10 in  \citealt{1997ApJS..108..261B}) to the south-west of S\,10, and another (smaller) shock-ionized region (the S\,9 SNR in \citealt{1997ApJS..108..261B}) at the southernmost end, without an X-ray counterpart. The alignment of X-ray knots and shock-ionized optical line emission in S\,10 is strongly suggestive of a jet.}
  \label{xray_vlt_im}
   \vspace{-0.2cm}
\end{figure}

\begin{figure}
\centering
\includegraphics[width=0.52\textwidth]{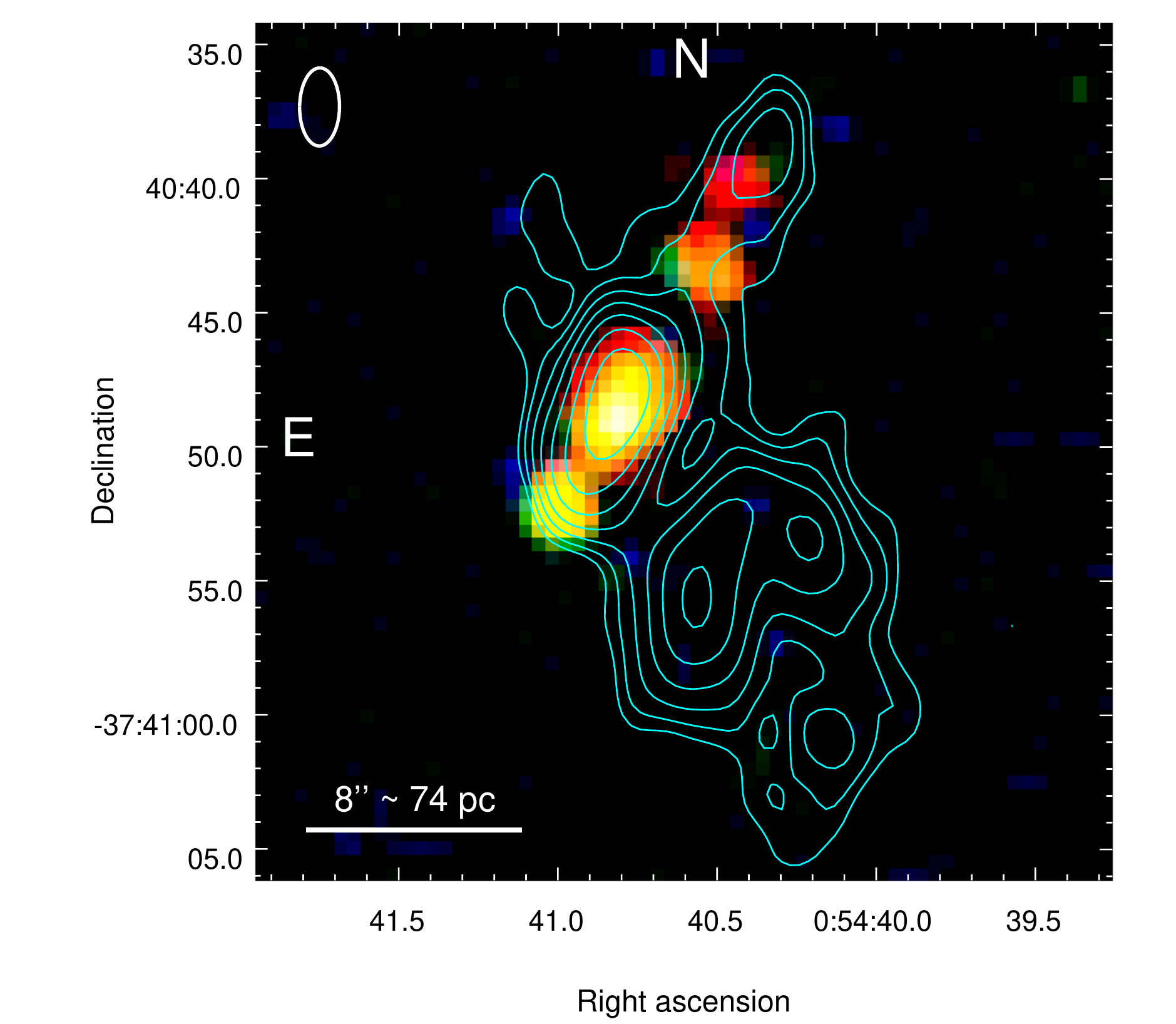}
\includegraphics[width=0.52\textwidth]{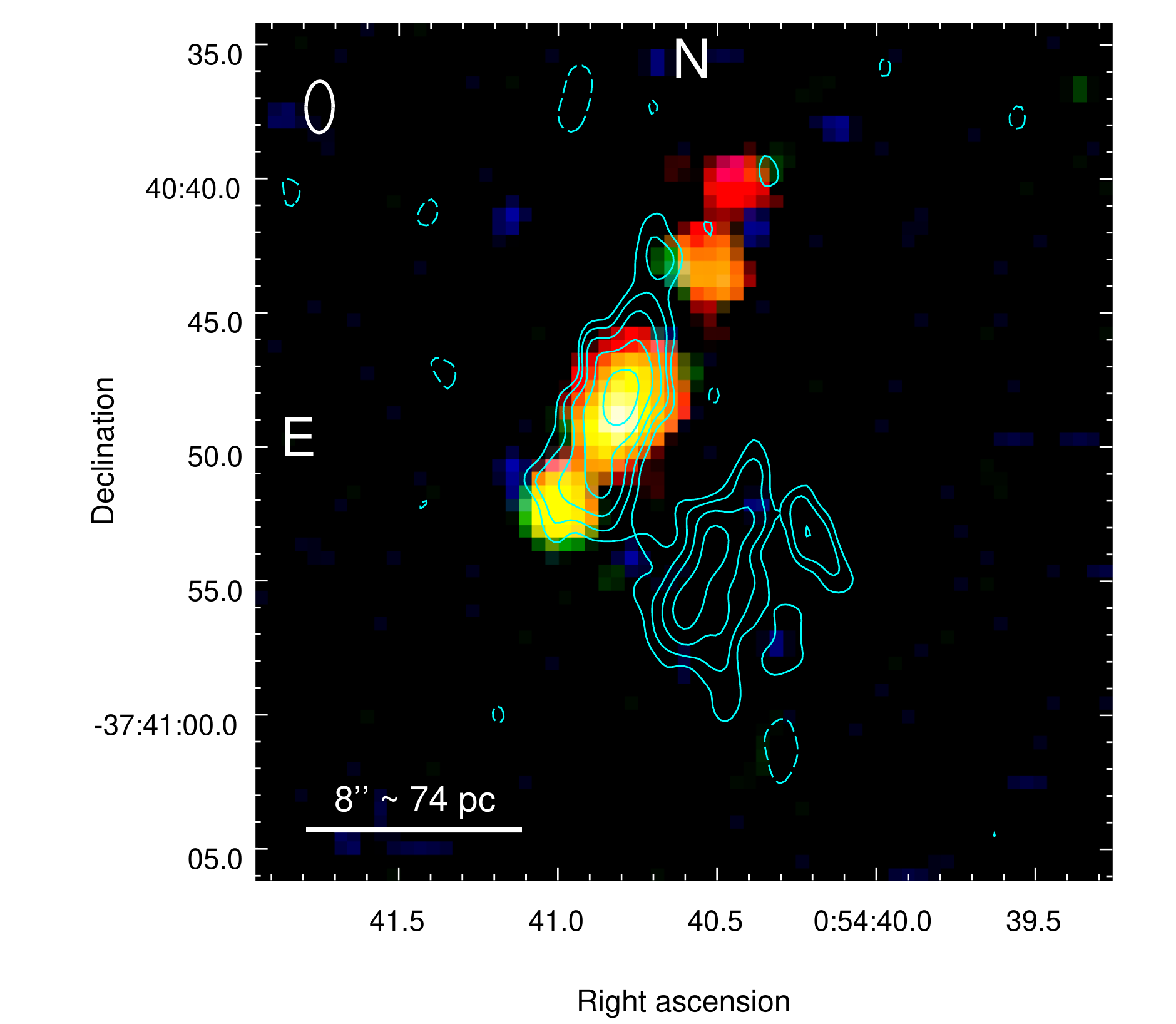}
 \caption{Top panel: Stacked {\it Chandra} ACIS-I image of NGC\,300 S\,10. Red represents 0.3--1.0 keV, green 1--2 keV, and blue 2--7 keV. The image has been adaptively smoothed with a Gaussian kernel of $\sigma=1$ pixel and has a pixel size of 0\barcs492. Cyan contours show the ATCA 5.5\,GHz map; contour levels are $2\sigma\times\sqrt{2}^{n}$, where $\sigma = 2.72\micro\jansky$\perbeam. The beam size is 2\barcs91 $\times$ 1\barcs47. Similarly to the H$\alpha$ emission (Figure \ref{xray_vlt_im}), the radio emission consists of a component associated and aligned with the X-ray knots in the S\,10 region, plus a component associated with the H\,10 H II region \citep{1997ApJS..108..261B} and another component corresponding to the S\,9 SNR \citep{1997ApJS..108..261B}. Bottom panel: Same as top panel, but the contours are now from the ATCA 9\,GHz map; contour levels are $2\sigma\times\sqrt{2}^{n}$, where $\sigma = 2.78\micro\jansky$\perbeam. The beam size is 1\barcs91 $\times$ 1\barcs01. The radio emission at S\,10 is spatially resolved both along the direction of the jet, and in the transverse direction, both at 5.5 GHz and at 9 GHz.}
  \label{xray_im}
\end{figure}

Searching for additional S\,26/SS\,433-type microquasars among supernova remnant candidates in nearby galaxies we noticed an intriguing source previously identified as a bright optical SNR by \cite{1980A&AS...40...67D} (listed as source number 2 in their Table 3; see also the finding chart in their Fig.~3). It is located in the nearby, late-type spiral NGC\,300, at a distance of $1.88 \pm 0.05$ Mpc \citep{2005ApJ...628..695G}. The same optical source was observed and studied in more detail by \cite{1997ApJS..108..261B}, who list it as NGC\,300 S\,10 = DDB2. In both studies, the SNR identification is based on the high ratio between [S {\sc ii}]$\lambda\lambda$6716,6732 and H$\alpha$ line emission; line ratios [S {\sc ii}]:H$\alpha$ $\gtrsim 0.4$ are indicative of shock-ionized gas \citep{1973ApJ...180..725M,1978A&A....63...63D}. What is striking about S\,10 is that we found four X-ray sources spatially resolved by {\it Chandra}, aligned in the plane of the sky and associated with the shock-ionized H$\alpha$ emission (Figure \ref{xray_vlt_im}). This is very unusual for an SNR; instead, we interpret the X-ray appearance as an unambiguous signature of a jet. Henceforth, we refer to those four X-ray sources as knots 1 through 4 (Figure \ref{xray_vlt_im}). Spurred by this discovery, we then observed the field in the radio band with the Australia Telescope Compact Array (ATCA). Previous studies reported associated radio emission \citep{2000ApJ...544..780P, 2004A&A...425..443P}. With our new ATCA data, we found a bright, elongated radio nebula, overlapping with the X-ray jet (Figure \ref{xray_im}). This is further evidence for the presence of collimated, relativistic ejections.

In this paper, we report on those discoveries, and analyse the multiband properties of S\,10, using a combination of archival and new data to probe the connection between the X-ray jet, radio nebula and accretion phases of the central engine. In Section \ref{sc:data_analysis} we outline our data reduction techniques; in Section \ref{sc:results} we present our X-ray, radio and optical results; and in Section \ref{sc:discussion} we discuss the energetics of S\,10 and compare it to super- and sub-Eddington jet sources.

\vspace{-0.6cm}

\section{Data Analysis} \label{sc:data_analysis}

\subsection{X-ray observations}

NGC\,300 has been observed by {\it Chandra} a total of five times. However, in one of those observations (ID 9883) the candidate microquasar target of our study does not fall on any of the chips. A second short observation (ID 7072), taken with HRC-I, does not have the sensitivity required to detect the source. Thus, we only used three of the five observations for our X-ray data analysis (Table \ref{obs_tab}): ACIS-I observations 12238, 16028 and 16029. We downloaded the data from the public archive, and reprocessed them using standard tasks from the Chandra Interactive Analysis of Observations ({\sc ciao}) Version 4.9 software package \citep{2006SPIE.6270E..1VF}. We filtered out high particle background intervals. For our imaging analysis, we used HEASARC's {\sc {DS9}} visualisation package. After we identified a number of discrete X-ray sources associated with the target of our study (as discussed in Section \ref{xray_results_sec}), we used the {\sc ciao} task {\it specextract} to extract the background-subtracted spectrum for each source, in each observation. For the two southernmost sources (knots 1 and 2), we extracted the source counts from circular regions of radius $3^{\prime\prime}$; for the other two sources (knots 3 and 4), we used elliptical regions of axes $3^{\prime\prime}\times2^{\prime\prime}$ (position angle $=330^{\circ}$), to avoid contamination from the brighter, neighbouring sources. A local background region was selected, approximately 3 times larger than the source regions. Spectral fitting was performed using {\small XSPEC} version 12.9.1 \citep{1996ASPC..101...17A}. Because of the low number of counts for each source, to test the goodness of our fits we used {\small XSPEC}'s implementation of W-statistics, which is Cash statistics \citep{1979ApJ...228..939C} modified for a background-subtracted spectrum.


\begin{table}
    \centering
    \caption{Observations used for this work. 
    The exposure time for the radio observations refers to the total time on source.}    
    \begin{tabular*}{0.48\textwidth}{l @{\extracolsep{\fill}} ccc}
        \hline\hline
        Telescope & Obs ID/Filter/$\nu$ & Obs Date & Exposure \\\hline
        {\it Chandra}/ACIS-I & 12238 & 2010-09-24 & 63.00\,\kilo\second \\
        & 16028 & 2014-05-16 & 64.24\,\kilo\second \\
        & 16029 & 2014-11-17 & 61.27\,\kilo\second \\
        & Total & & 188.51\,\kilo\second  \\\hline
        {\it HST}/ACS & F814W & 2006-11-09 & 1542\,\second \\
        & F606W & 2006-11-09 & 1515\,\second \\
        & F475W & 2006-11-09  & 1488\,\second \\\hline
        VLT/FORS2 & H$\alpha$ & 2010-07-10  & 180\,\second$^*$ \\
        & H$\alpha$/4500 & 2010-07-10  & 180\,\second$^*$ \\\hline
        ATCA & 5.5/9\,\GHz & 2015-10-21 & 4.53\,\hour \\
        & 5.5/9.0\,\GHz & 2015-10-22 & 10.75\,\hour \\
        & 5.5/9.0\,\GHz & 2015-10-23 & 10.32\,\hour \\
        & 5.5/9.0\,\GHz & 2016-08-25 & 9.95\,\hour \\
        & 5.5/9.0\,\GHz & 2016-08-26 & 9.86\,\hour \\
        & Total & & 45.41\,\hour\\
        \hline
    \end{tabular*}
    \begin{flushleft}
    $^*$ Stack of two 90\,s exposures\\
    \end{flushleft}
    \label{obs_tab}
    \vspace{-0.35cm}
\end{table}

\begin{figure*}
\centering
\includegraphics[height=0.4\textwidth]{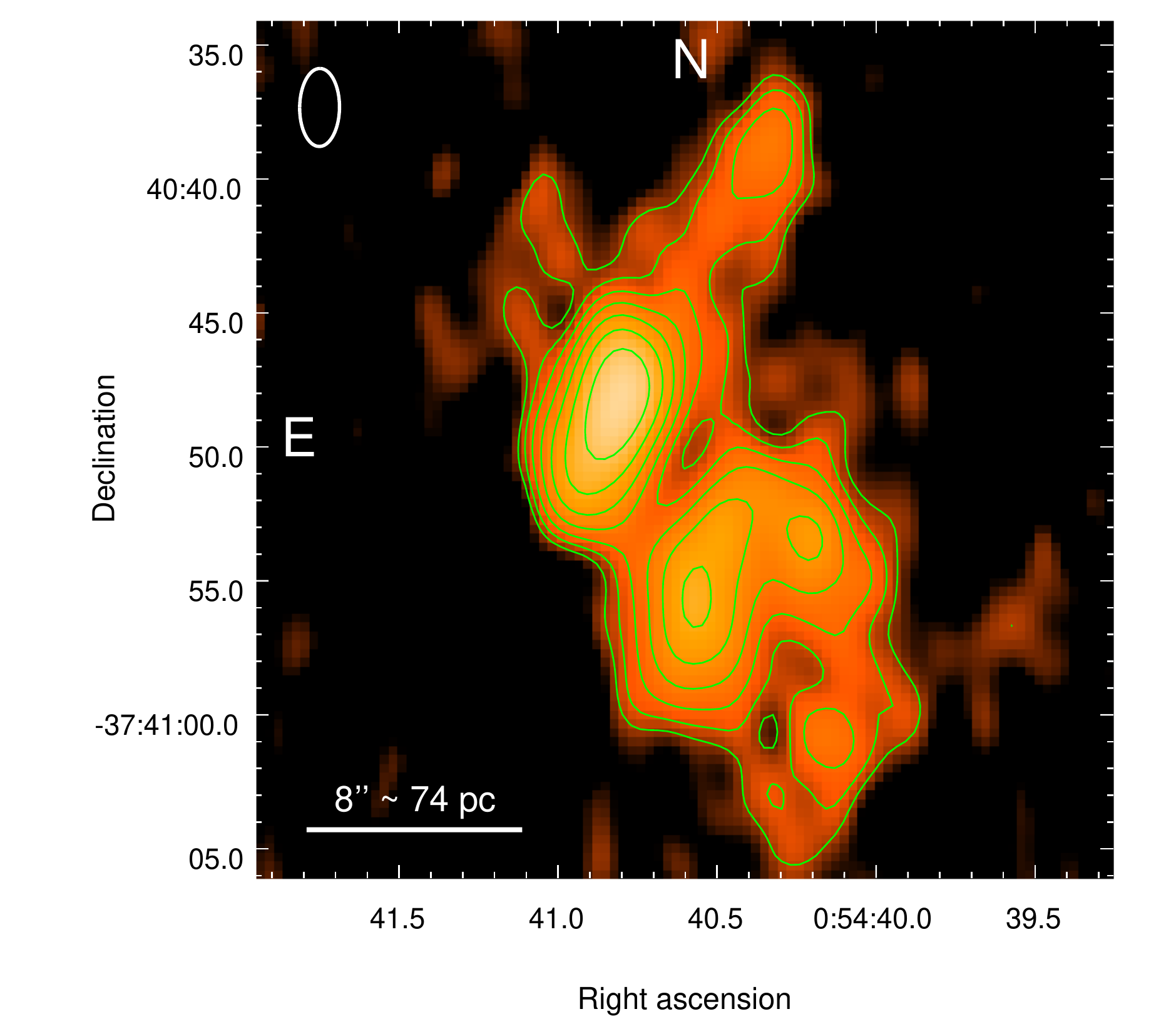}
\includegraphics[height=0.403\textwidth]{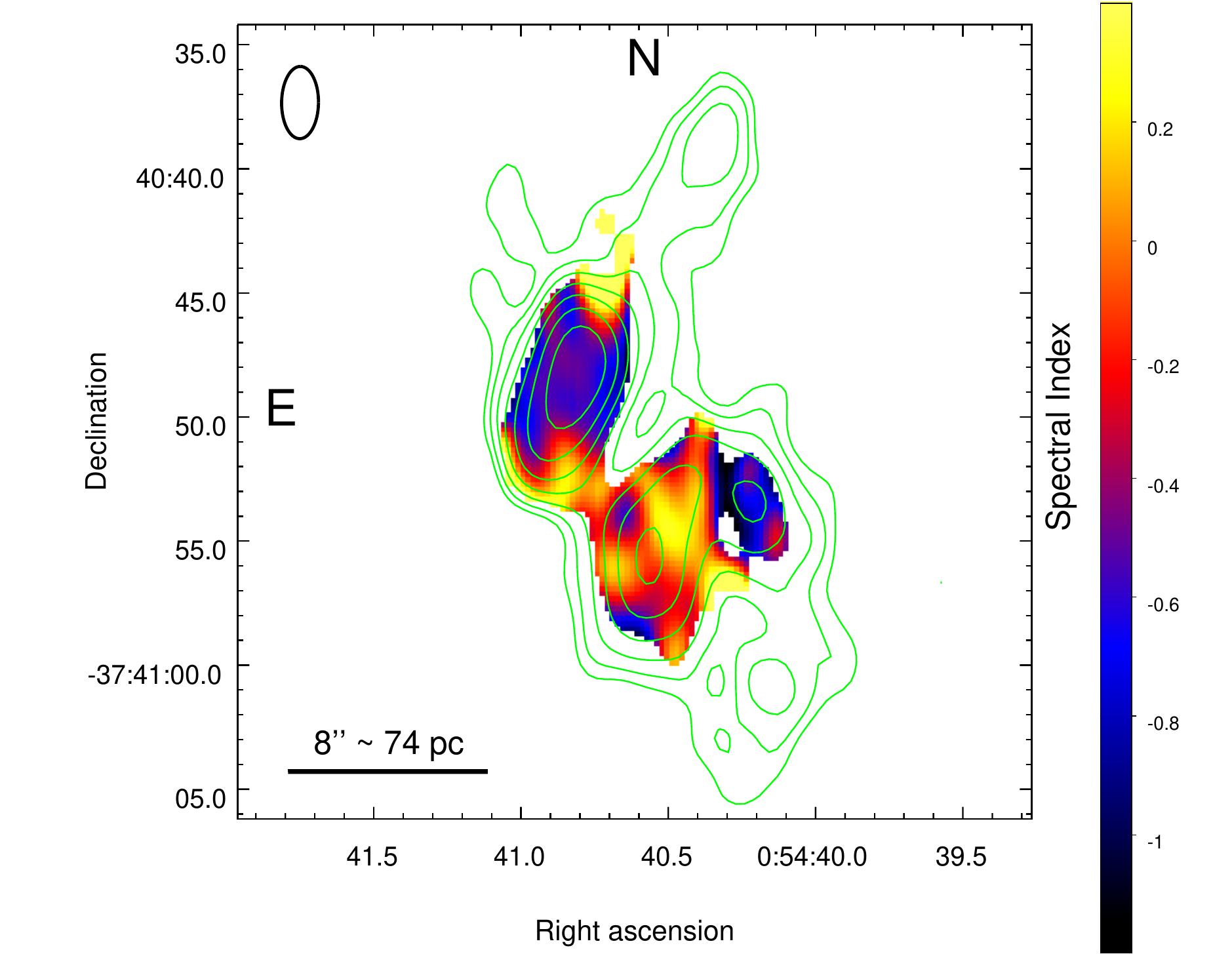}\\
\includegraphics[height=0.4\textwidth]{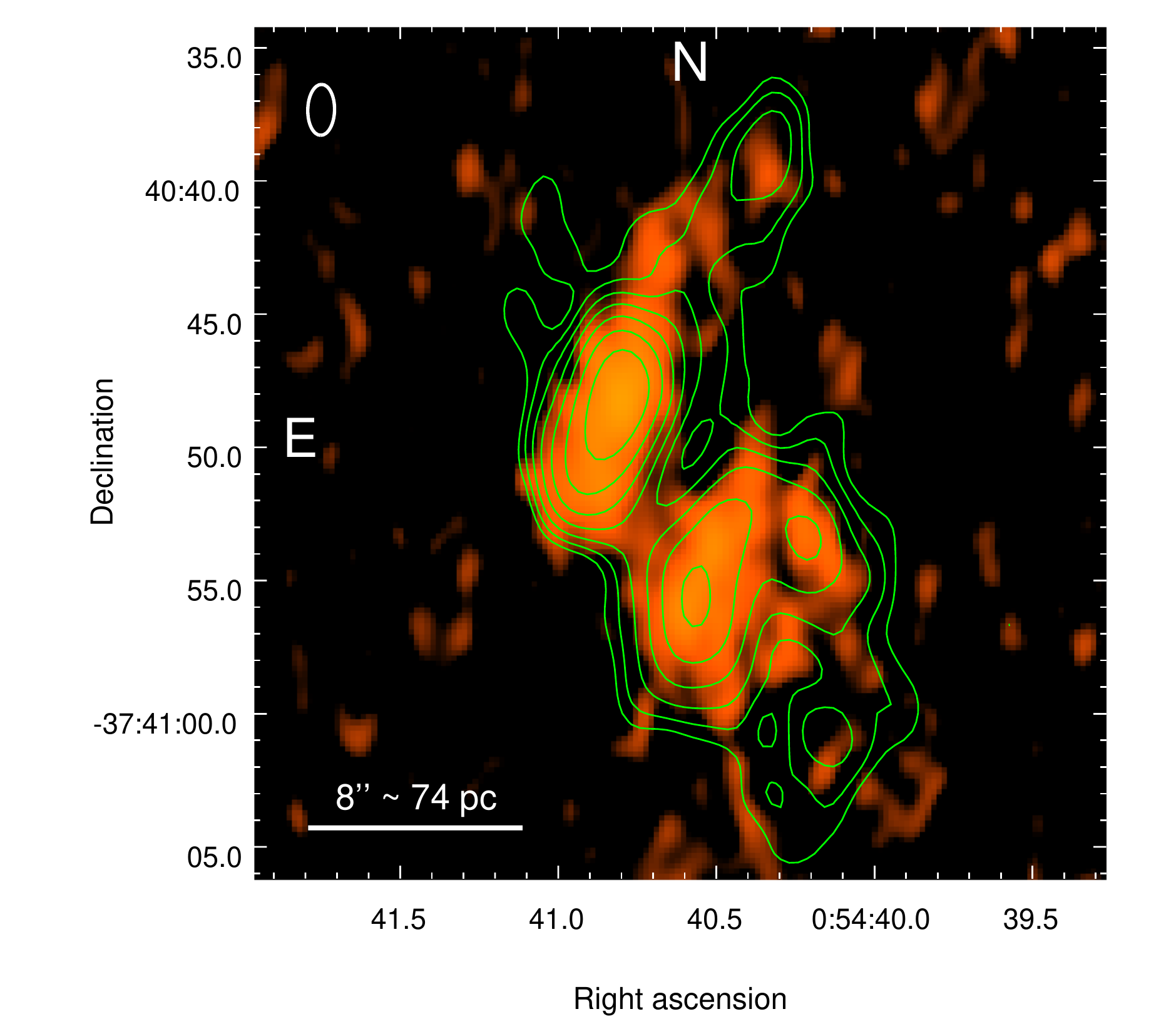}
\includegraphics[height=0.403\textwidth]{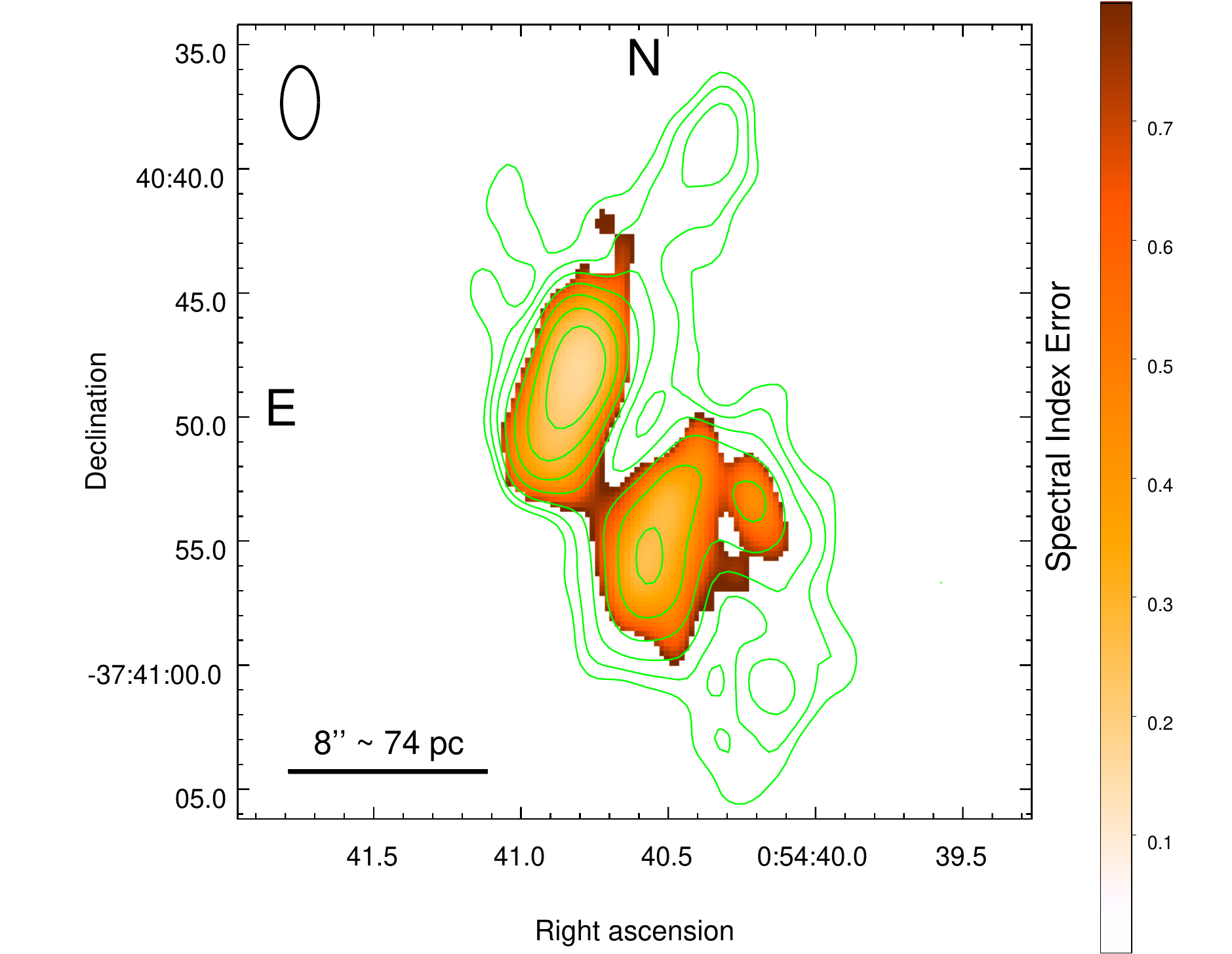}
 \caption{Top left panel: ATCA 5.5\,GHz image. Contours levels are $2\sigma\times\sqrt{2}^{n}$, where $\sigma = 2.72\micro\jansky$\perbeam. Bottom left panel: ATCA 9\,GHz image with 5.5\,GHz contours for reference. In both images, the radio nebula is aligned with the X-ray knots, which  suggests that both are the result of a jet. Top right panel: Spectral index map made from the ATCA 5.5 and 9\,GHz images. The 9\,GHz image was convolved with the 5.5\,GHz beam. We mask values below 3$\sigma$ in their respective 5.5 and 9\,GHz images. Bottom right panel: Error values for our spectral index map. We find that the central source has a steep spectral index ($\alpha=-0.4\pm0.2$), consistent with optically-thin synchrotron emission. To the southwest we see the H{\sc ii} region, H\,10 \citep{1997ApJS..108..261B}, which, as expected, has a flatter spectral index ($\alpha\sim-0.1-0.3\pm0.3$) than S\,10.}
  \label{radio_im}
\end{figure*}

\subsection{Radio observations} \label{rad_dat_red_sec}

NGC\,300 S\,10 was observed with the ATCA over three consecutive nights, starting on 2015 October 21 (proposal ID C3050). A total integration time on source of 25.6 hours was achieved. We then re-observed NGC\,300 one year later, this time for two consecutive nights, 2016 August 25--26 (proposal ID C3120). For the 2015 observations, S\,10 was the target and hence at the phase centre, while the follow-up observations were offset to observe another source in the same field, with the phase centre at RA $= 00^{h}$55$^{m}$02\bsecs5, Dec $= -37^{\circ}$40$'$10\barcs5, approximately 4\bmins5 from S\,10. For both observation sets, two 2048\,MHz bands were observed simultaneously, centred at 5.5\,GHz and 9.0\,GHz, using the Compact Array Broadband Backend \citep[CABB;][]{2011MNRAS.416..832W}. The telescope was in its extended 6A configuration during all of the observations, resulting in a maximum baseline of 5938.8\,m. B1934$-$638 was used as both a bandpass and flux calibrator. We used B0104$-$408 as the phase calibrator. The bandpass/flux calibrator was observed at the start of the observation for 10 minutes while the phase calibrator was observed for 1 minute every 15-minute block.

We used the {\small MIRIAD} package \citep{1995ASPC...77..433S} for gain and phase calibration. The data were then imaged with the Common Astronomy Software Application \citep[CASA;][]{2007ASPC..376..127M} software package, with the {\small CLEAN} algorithm. All three nights (two nights for the 2016 observing run) of observations were stacked for each frequency and we imaged the data with Briggs weighting at a robust value of one. Finally, we combined the two observing runs to create a single mosaic image for each frequency. The cleaned, primary-beam-corrected images for both frequencies are shown in Figure \ref{radio_im}. The 5.5\,GHz image has a Gaussian restoring beam of 2\barcs91 $\times$ 1\barcs47 (position angle $-0\bdegs6$ East of North) and an rms noise level of $2.72\,\mu$\jansky\perbeam at the location of S\,10. The 9\,GHz image has a Gaussian restoring beam of 1\barcs91 $\times$ 1\barcs01 (position angle $-1\bdegs2$ East of North) and a local rms noise level of $2.78\,\mu$\jansky\perbeam.


In order to construct a spectral index map (Figure \ref{radio_im}, right), we first convolved the 9\,GHz image with the beam of the 5.5\,GHz image. The 5.5\,GHz image was then re-gridded to match the 9\,GHz image, with the CASA task {\sc imregrid}. We then created a two-point spectral index map (Figure \ref{radio_im}, top right), along with a corresponding error map (Figure \ref{radio_im}, bottom right). Both data sets were masked to their respective 3$\sigma$ thresholds prior to use.

\vspace{-0.3cm}
\subsection{Optical observations}


\subsubsection{HST imaging}

\begin{figure}
\hspace{-0.5cm}
\includegraphics[width=0.52\textwidth]{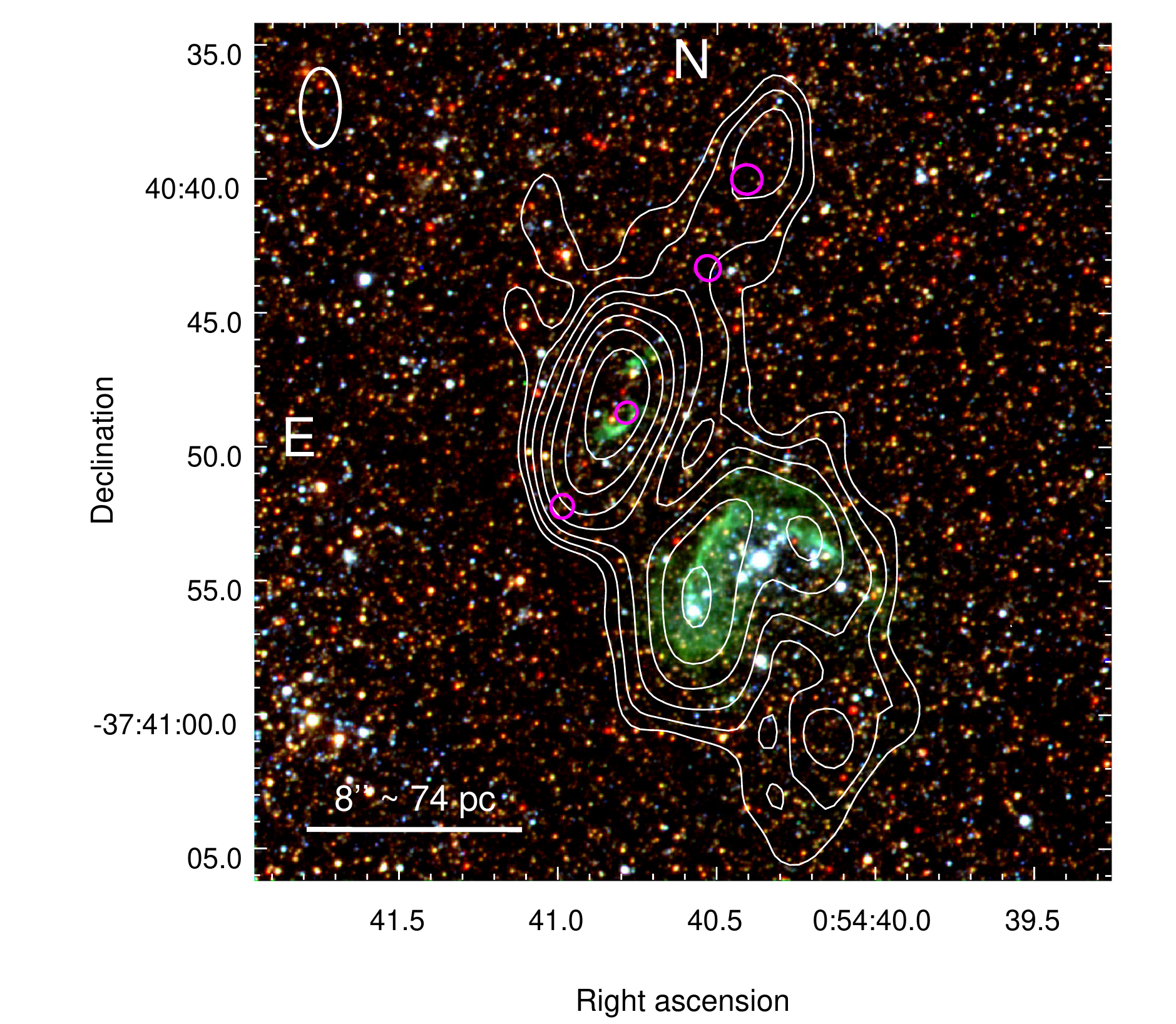}\\
\vspace{-0.1cm}
\hspace{-0.5cm}
\includegraphics[width=0.52\textwidth]{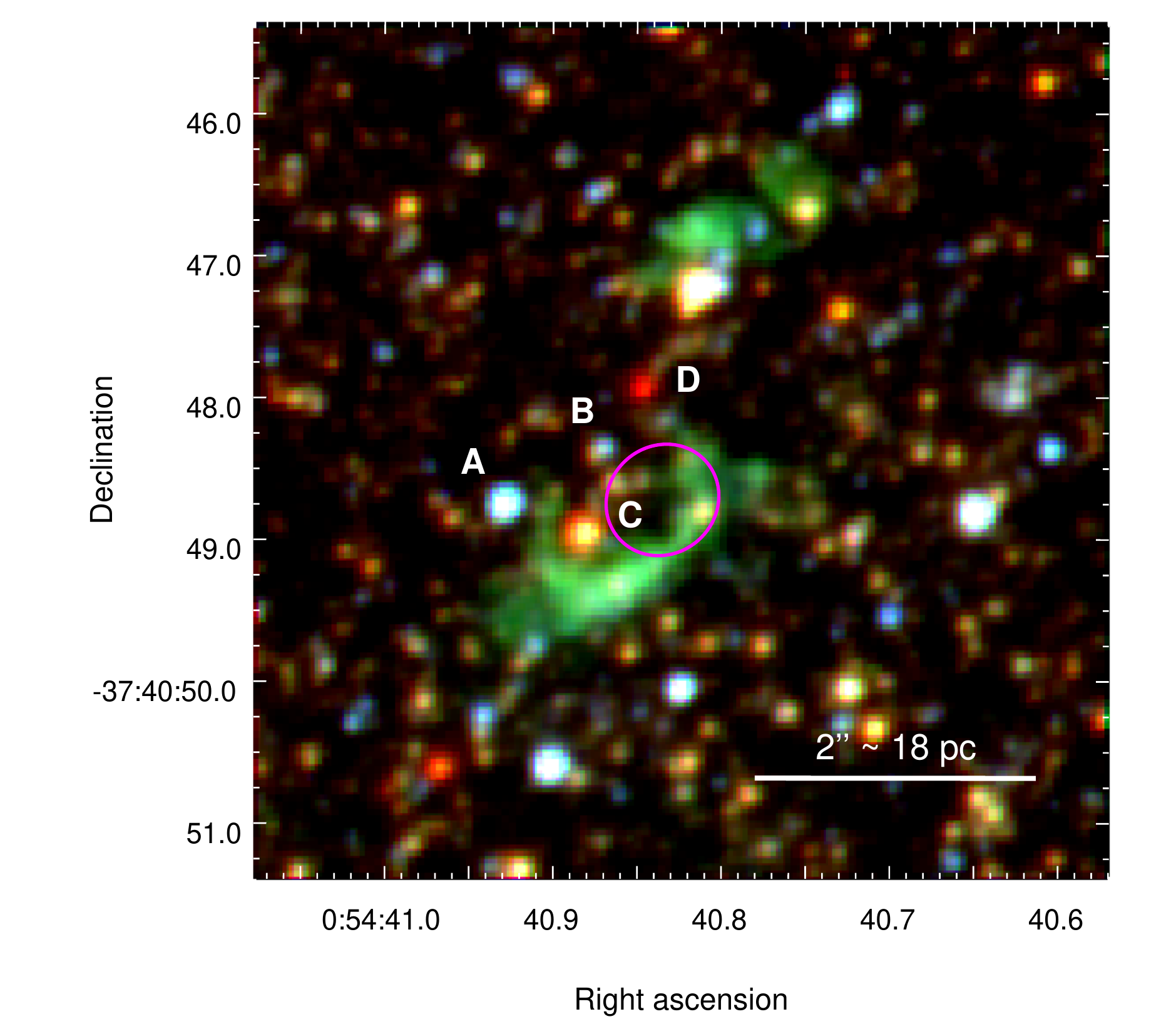}
\vspace{-0.1cm}
 \caption{Top panel: {\it HST}/ACS-WFC RGB colour image of S\,10. Red represents the F814W band, green is the F606W band and blue is the F475W band. Overlaid are the ATCA 5.5\,GHz  contours in white and, in magenta, the 95\% confidence {\it Chandra} error circles (Section \ref{sc:astrometry}) for the peak-emission location of each knot. There is nebular emission (likely H$\alpha$ plus [N II], which are covered by the F606W bandpass) coincident with X-ray knot 2 and the peak radio emission. The star-forming region and photoionized nebula H\,10 \citep{1997ApJS..108..261B} is to the south-west. Bottom panel: Zoomed-in image of the inner part of the S\,10 complex around the extended X-ray knot 2, more clearly showing the two separate nebulae (in green). The magenta circle represents the 95\% confidence {\it Chandra} central position of the brightest knot (Section \ref{sc:stellar_count}). Letters indicate potential optical counterparts for the accreting compact object, origin of the jet. 
 The distinct morphology of the line-emitting nebula around the peak X-ray and radio emission could be a result of anisotropic ejecta from a progenitor SN or it could be gas shock-ionized by the jet and the counterjet.}
  \label{opt_im}
\end{figure}

We downloaded publicly available {\it Hubble Space Telescope} ({\it HST}) data for NGC\,300 from the Hubble Legacy Archive\footnote{http://hla.stsci.edu/hlaview.html}. The observations (proposal ID 10915) were taken on 2006 November 9 with the Wide Field Channel (WFC) of the Advanced Camera for Surveys (ACS). We used the broadband filters F475W, F606W and F814W, with exposure times 1488$\,\second$, 1515$\,\second$ and 1542$\,\second$ respectively. We created a three-colour image using the F475W, F606W and F814W bands (Figure \ref{opt_im}).

We performed aperture photometry on potential optical counterparts to the X-ray knots. Source count rates for each filter were extracted from $0.15$ arcsec (3 pixels) circular regions. We selected nearby backgrounds at least three times larger than the source regions. We used the encircled energy fractions of \citet{2005PASP..117.1049S} to determine the aperture correction and calculate the source count rates at infinite aperture. Finally, we converted count rates to physical magnitudes using the zeropoint tables for ACS-WFC\footnote{http://www.stsci.edu/hst/acs/analysis/zeropoints}.



\subsubsection{VLT imaging}

Finally, from the ESO Science Archive Facility\footnote{http://archive.eso.org/cms.html}, we downloaded publicly available Very Large Telescope (VLT) data to better study the narrow-band H$\alpha$ emission. NGC\,300 S\,10 was observed on 2010 July 10 with two consecutive FORS2 90\,s exposures (with the first exposure starting at Universal Time 09:54:45) with the ``H\_Alpha+83'' interference filter\footnote{http://www.eso.org/sci/facilities/paranal/instruments/fors/inst/ Filters.html}. The filter has a central wavelength of 6563\,\AA\ and a full width at half maximum (FWHM) of 61\,\AA; thus, it also includes the [N {\sc ii}]$\lambda \lambda$6548,6583 lines. The collimator was in standard resolution. The seeing was $\approx$0\barcs6 and the airmass 1.03. This was followed by two consecutive 90\,s exposures with the ``H\_Alpha/4500+61'' interference filter, centred at 6665\,\AA, for continuum subtraction.

\begin{figure}
\hspace{-0.5cm}
\includegraphics[width=0.52\textwidth]{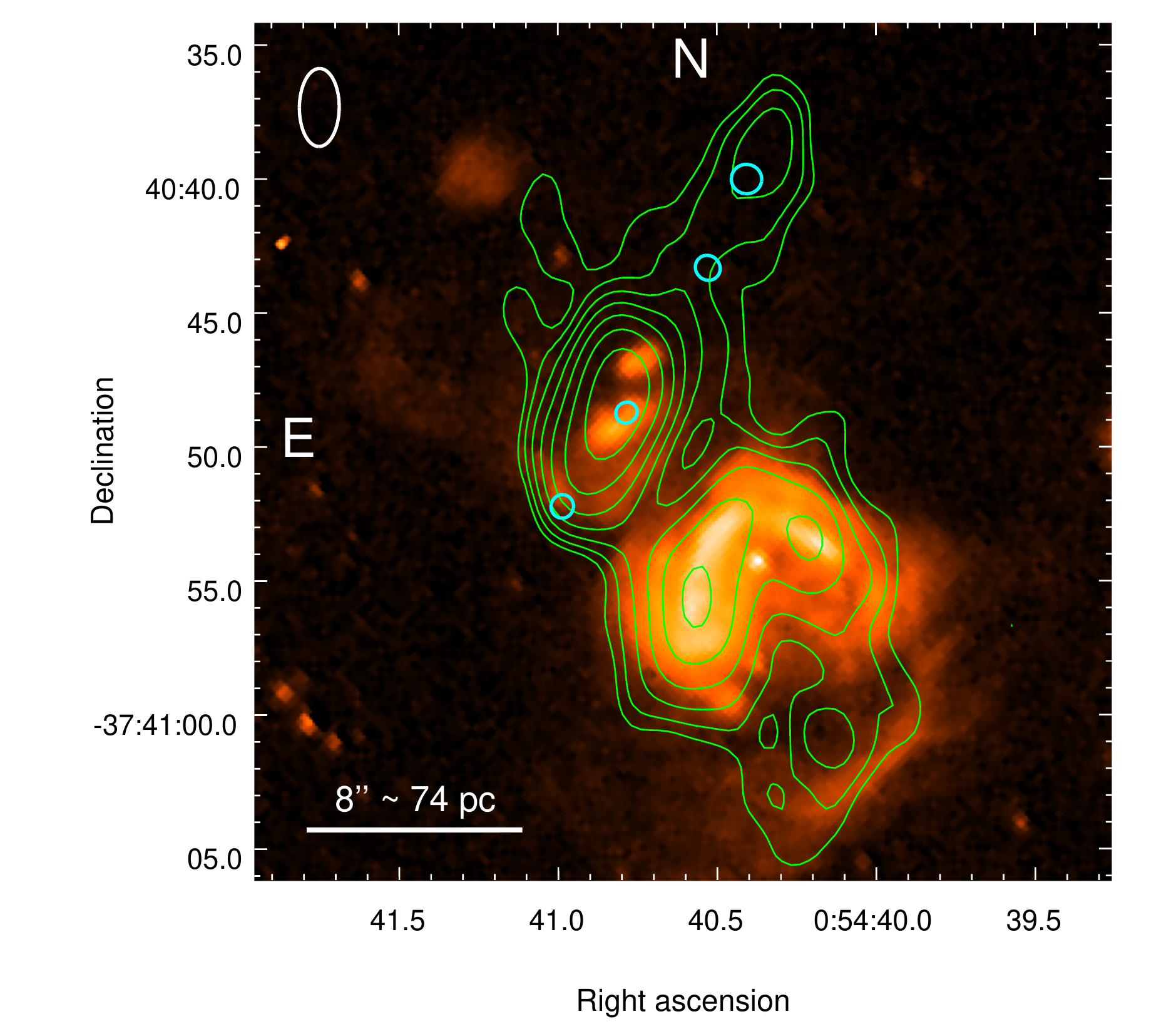}\\
\vspace{-0.3cm}
\hspace{-0.5cm}
\includegraphics[width=0.52\textwidth]{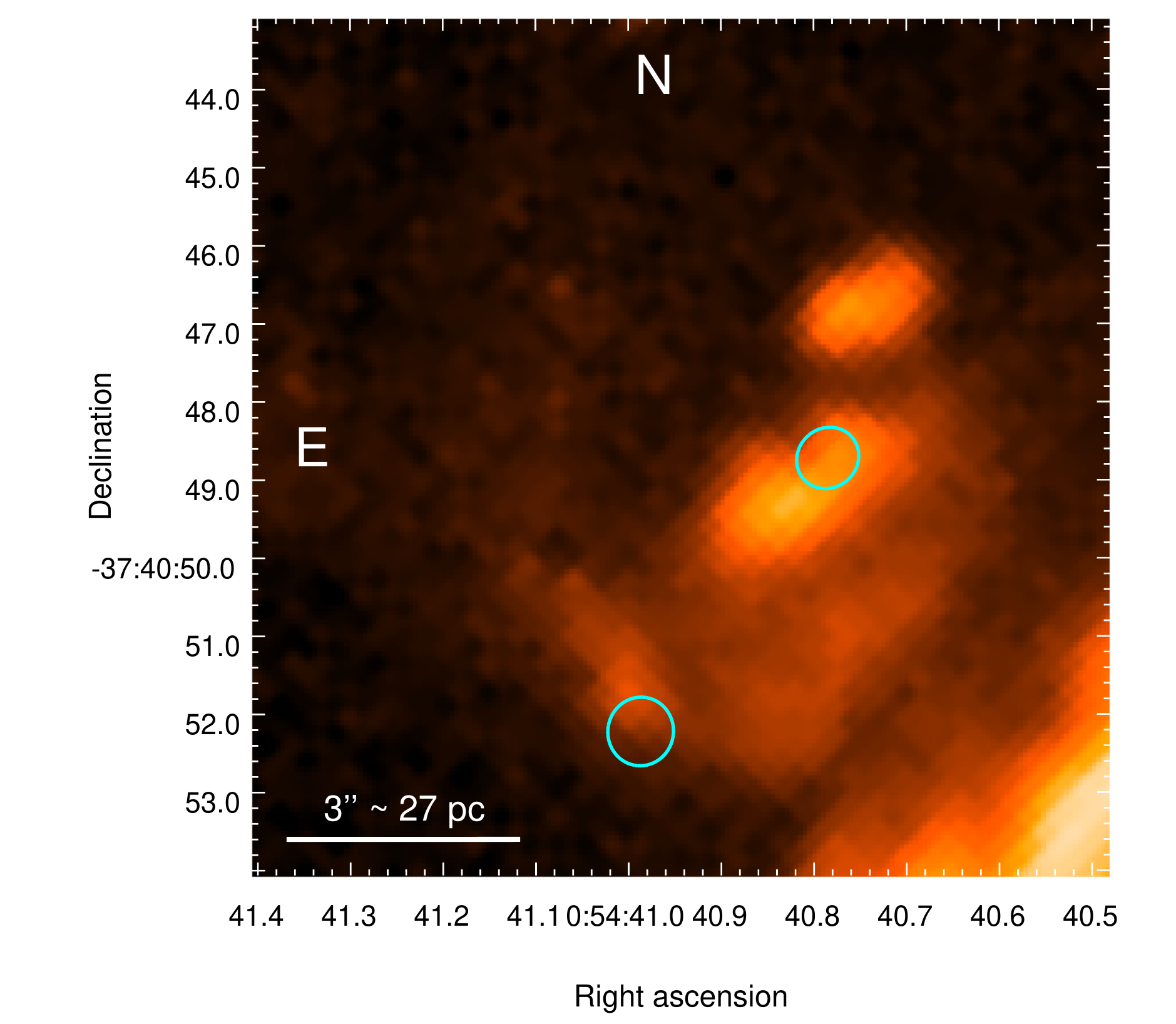}
\vspace{-0.1cm}
 \caption{Top panel: smoothed VLT/FORS2 continuum-subtracted H$\alpha$ image of S\,10 with the {\it Chandra} 95\% error circles for the four X-ray knots in cyan and 5.5\,GHz ATCA contours overlaid in green. As already noted in the {\it HST} image (Figure \ref{opt_im}), the two regions with the strongest H$\alpha$ emission are located on either side of X-ray knot 2. However, the VLT image also shows a larger H$\alpha$ nebula, extending south as far as X-ray knot 1. Bottom: Zoomed-in view of the core region, from the continuum-subtracted VLT/FORS2 image. The cyan circles mark the position of X-ray knots 1 and 2.}
  \label{vlt_im}
\end{figure}

We stacked the two 90\,s H$\alpha$ exposures and the corresponding continuum exposures using the Image Reduction and Analysis ({\sc iraf}) software Version 2.16 \citep{1993ASPC...52..173T} package {\it imcombine}. 
We aligned the stacked H$\alpha$ and continuum images with the {\sc iraf} package {\it ccmap}, relying on $\approx30$ common bright sources. Finally, we subtracted the continuum using the {\sc iraf} package {\it imarith}. The resulting image is displayed in Figure \ref{vlt_im}. We used the continuum-subtracted image for flux measurements (Section \ref{sc:optical_results}).

\subsection{Astrometric alignment} \label{sc:astrometry}

The default {\it Chandra}/ACIS-I astrometry is known to be accurate within $\approx$0\barcs75 for 90\% of the observations\footnote{http://cxc.harvard.edu/cal/ASPECT/celmon/}. To improve on this value, we aligned the stacked {\it Chandra} image to the DR1 {\it Gaia} astrometric frame  \citep{2016A&A...595A...4L,2016A&A...595A...2G,2016A&A...595A...1G}, which has a mean uncertainty $\sigma \approx0.3$ mas. Because of the low number of X-ray sources, we were only able to find five common, bright point sources suitable for frame alignment. After alignment, we find position error residuals of $\approx$0\barcs1--0\barcs2.  We determined the central position of each X-ray source using the {\sc ciao} task {\it wavdetect} on the stacked {\it Chandra}/ACIS-I image. To estimate their respective confidence error circles, we calculated the 95\% statistical uncertainty for the central position of each source using Eq.~(5) from \citet{2005ApJ...635..907H}; we then combined (in quadrature) those statistical uncertainties and the {\it Chandra}/{\it Gaia} alignment residuals. We also aligned our {\it HST} images in the F475W, F606W and F814W bands and our VLT H$\alpha$/4500 and H$\alpha$ bands onto the {\it Gaia} astrometric frame, using $>100$ common, bright, non-saturated sources. This provides a positional accuracy within 0\barcs009 for the {\it HST} images and 0\barcs03 for the VLT images. We assume that the radio and {\it Gaia} astrometric frames are consistent with each other, within the relative uncertainties.



\begin{table}
    \centering
    \caption{Central position of each X-ray knot; knot 2 is spatially resolved, while knots 1, 3 and 4 are unresolved.}
    \begin{tabular*}{0.48\textwidth}{l @{\extracolsep{\fill}} cc}
        \hline\hline
        & RA (J2000) & Dec (J2000)\\\hline\\[-6pt]
        Knot 1 & 00$^h$54$^m$40\bsecs99 $(\pm0\barcs44)$ & $-37^{\circ}$40$'$52\barcs2 $(\pm0\barcs42)$ \\
        Knot 2 & 00$^h$54$^m$40\bsecs79 $(\pm0\barcs41)$ & $-37^{\circ}$40$'$48\barcs7 $(\pm0\barcs38)$ \\
        Knot 3 & 00$^h$54$^m$40\bsecs53 $(\pm0\barcs48)$ & $-37^{\circ}$40$'$43\barcs3 $(\pm0\barcs45)$ \\
        Knot 4 & 00$^h$54$^m$40\bsecs41 $(\pm0\barcs57)$ & $-37^{\circ}$40$'$40\barcs0 $(\pm0\barcs55)$ \\
        \hline
    \end{tabular*}
    \label{coords_tab}
\end{table}

\section{Results} \label{sc:results}

\subsection{X-ray results} \label{xray_results_sec}

\subsubsection{X-ray knots} \label{xray_knots_results_sec}

The {\it Chandra} data reveal the presence of four X-ray sources that appear linearly aligned with a total projected length of $\approx17^{\prime\prime}\approx150\,\parsec$ (Figure \ref{xray_im}). No other X-ray sources are detected within a circle of radius $\approx1'$. Additionally, we only expect $\approx4\E{-3}$ extragalactic sources \citep{2012ApJ...752...46L} with $0.5-2.0\,\kev$ fluxes greater than $f_{0.5-2.0\,\kev}=1.9\E{-15}\,\flux$ (the flux of the dimmest source, knot 4) within a region equivalent to the projected area of the radio bubble (i.e.,\,  $19'' \times 6''\approx8.80\E{-6}$ deg$^2$); therefore a physical connection between those aligned sources is highly likely. We shall refer to those four visually distinct X-ray sources as knots 1 to 4 (starting from the southernmost one). The central positions of those four knots are listed in Table \ref{coords_tab}.

Of those four sources, knot 2 appears, at first sight, spatially extended (Figure \ref{xray_im}) and larger than the other three. We quantified this visual impression with the {\sc{MARX}}\footnote{http://cxc.harvard.edu/ciao/threads/marx/} simulation software. Using the task {\it simulate\_psf}, we simulated 1000 point-spread functions (PSFs) at the location of S\,10 on the ACIS-I chip during ObsID 16028. For this ensemble of simulated PSFs, the average encircled energy fraction as a function of radius was determined with the {\sc ciao} task {\it ecf\_calc}. We compared this value to the observed encircled energy fraction of knot 2 during ObsID 16028. The results are displayed in Figure \ref{psf_im}. The shape of the PSF does not match that observed for knot 2 and proves that knot 2 is indeed extended. It is consistent with two point-like sources separated by $\approx$1\barcs2 and oriented roughly along the same direction defined by the other knots.

There are insufficient counts to do meaningful spectral analysis on the individual X-ray knots for individual epochs. However, we can investigate source variability and evolution of the hardness ratios from the three epochs spaced over $\approx$4 years. For the hardness ratios, we defined the soft band as $0.3$--$1.2\,\kev$ (by analogy with the combined ultra-soft and soft bands in the Chandra Source Catalogue; \citealt{2010ApJS..189...37E}), and $1.2$--$7.0\,\kev$ for the hard band. 
Since the ACIS-I detector has lost sensitivity over the years, especially in the soft band, we must correct the count rates measured in ObsID 12338 (Cycle 11) to make them comparable with the count rates measured in Cycle 15 (ObsIDs 16028 and 16029). We did that using the online tool {\sc pimms}\footnote{http://cxc.harvard.edu/toolkit/pimms.jsp} version 4.8e. For knot 1, we assumed a thermal plasma model with temperature $kT\approx0.8\,\kev$, while for knots 2, 3 and 4 we assumed a temperature of $kT\approx0.4\,\kev$. These assumptions are based on our spectral fitting after we combined the spectra from all three {\it Chandra} epochs, as described later in this section\footnote{The cycle-to-cycle corrections to the ACIS-I count rates are relatively small and do not substantially depend on the choice of model; the results listed in Table \ref{rates_tab} are essentially unchanged (differences $\lesssim10\%$) if we assume the same plasma temperature for all knots, or if we use a steep power-law as spectral model.}.
The cycle-corrected hard and soft count rates are displayed in Table \ref{rates_tab}. We find no significant variability for any of the knots, in either band, over all three epochs. The corresponding hardness ratios are displayed in Table \ref{hr_tab}. While knots 2, 3 and 4 all have similar spectral hardnesses, knot 1 appears harder. As with the count rates, no significant variability is detected across all four knots and all three epochs.

In order to conduct spectral modelling of the knots, we stacked the three {\it Chandra} epochs. Each of the knots has insufficient counts for $\chi^2$ statistics and so w-statistics were used. We first attempted to fit each of the knots with a simple absorbed power-law model ({\it TBabs $\times$ po}). However all power-law fits yielded unphysically high absorption and photon indices. For example, fitting knot 2 with an absorbed power-law model yields a best-fit photon index $\Gamma \approx 10$ and column density $n_{\rm H}=10^{22}\,\centi\meter^{-2}$, for a w-statistic of 87.7 and 72 degrees of freedom. Instead, we find (Table \ref{xray_tab}) that all four knots are well described by thermal plasma models, such as {\it mekal} \citep{1985A&AS...62..197M,1986A&AS...65..511M} (Figure \ref{xray_spectra}). No additional intrinsic absorption is required, and thus we use a single {\it TBabs} component fixed to the line-of-sight column density towards NGC\,300 ($N_{\rm H} = 3\times10^{20}\,\centi\meter^{-2}$; \citealt{1990ARA&A..28..215D, 2005A&A...440..775K}). As expected from the hardness ratios (Table \ref{hr_tab}), knot 1 appears hotter than the rest. The unabsorbed X-ray luminosities of the knots ranges from $\approx$1 $\times 10^{36}\,\ergs$ for the dimmest knot (knot 4) to $\approx$7 $\times 10^{36}\,\ergs$ for the brightest (knot 2). For the spectrum of the combined emission of all knots, we find a best-fitting thermal plasma temperature $T\approx0.6\,\kev$ and an unabsorbed bolometric luminosity $L \approx 1.1 \times 10^{37}\,\ergs$. The model for this best-fitting thermal plasma is shown in Figure \ref{xray_spectra} with the fit parameters displayed in Table \ref{xray_tab}. While the spectrum is well fitted by thermal plasma alone, we also try a thermal plasma plus power-law model. We fix the power-law photon index to $\Gamma=1.7$, typical of X-ray binaries with X-ray luminosities similar to the core of S\,10 (see Sections \ref{sc:xray_core} and \ref{sc:synch_vs_thermal} for more details). However, we find no statistically significant improvement to the fit, with a maximum power-law contribution of $L<2\E{36}\,\ergs$ at the 90\% confidence level. Thus, by Occam's razor, our preferred model is the simple best-fitting thermal plasma model.


The normalisation constant for the {\it mekal} component (at zero redshift) is defined as $K=10^{-14}\,\int (n_e\,n_{\rm H}\,dV)/(4\pi d^2) \approx 10^{-14}\,n_e^2\,V/(4\pi d^2)$, where $V$ is the volume of the emitting region, $n_e$ is the electron density, $n_{\rm H}$ is the nuclear density and $d$ is the distance. We approximate the X-ray structure as five spherical sources (two of them corresponding to knot 2) with radii similar to or smaller than the {\it Chandra} resolution, $\approx$0\barcs4 $\approx 1.1\times10^{19}\,\centi\meter$; this corresponds to a total volume $V \lesssim 3\times10^{58}\,\centi\meter^{3}$ for the emitting plasma region. Thus, for the best-fitting normalisation $K \approx 9 \times10^{-6}$ in our spectral model (Table \ref{xray_tab}), we estimate a lower limit to the density of the hot plasma component, $n_e\gtrsim4\,\centi\meter^{-3}$.


\begin{table*}
    \centering
    \caption{{\it Chandra}/ACIS-I count rates for the hard (1.2--7.0 keV) and soft (0.3--1.2 keV) bands. {\it Chandra} Obs 12238 (Cycle 11) has been scaled to match the {\it Chandra} sensitivity of the other observations (Cycle 15). Count rates are in units of $10^{-4}$ count s$^{-1}$.} 
    \begin{tabular*}{\textwidth}{l @{\extracolsep{\fill}} rrrrrrrrrr}     
    \hline\hline
        Obs ID & \multicolumn{2}{c}{Knot 1} & \multicolumn{2}{c}{Knot 2} & \multicolumn{2}{c}{Knot 3} & \multicolumn{2}{c}{Knot 4} & \multicolumn{2}{c}{Combined}\\
        & \multicolumn{1}{c}{Hard} & \multicolumn{1}{c}{Soft} & \multicolumn{1}{c}{Hard} & \multicolumn{1}{c}{Soft} & \multicolumn{1}{c}{Hard} & \multicolumn{1}{c}{Soft} & \multicolumn{1}{c}{Hard} & \multicolumn{1}{c}{Soft} & \multicolumn{1}{c}{Hard} & \multicolumn{1}{c}{Soft} \\
        \hline
        12238 & $1.8\pm0.6$ & $2.3\pm0.6$ & $2.7\pm0.7$ & $9.5\pm1.1$ & $\leq0.4$ & $2.4\pm0.5$ & $0.4\pm0.3$ & $0.7\pm0.3$ & $5.0\pm1.0$ & $15.6\pm1.4$ \\
        16028 & $2.4\pm0.6$ & $1.9\pm0.5$ & $2.5\pm0.7$ & $10.9\pm1.3$ & $0.5\pm0.4$ & $2.3\pm0.6$ & $\leq0.6$ & $1.4\pm0.5$ & $6.0\pm1.0$ & $17.0\pm1.6$ \\
        16029 & $1.5\pm0.5$ & $3.9\pm0.8$ & $1.9\pm0.6$ & $7.0\pm1.1$ & $0.7\pm0.4$ & $1.3\pm0.5$ & $0.4\pm0.1$ & $0.3\pm0.2$ & $4.6\pm0.9$ & $12.9\pm1.5$\\
        Total & $2.1\pm0.3$ & $3.0\pm0.4$ & $2.4\pm0.4$ & $10.2\pm0.7$ & $0.5\pm0.2$ & $2.4\pm0.4$ & $0.3\pm0.2$ & $0.9\pm0.2$ & $5.2\pm0.6$ & $16.7\pm0.9$ \\
        \hline
    \end{tabular*}
    \label{rates_tab}
\end{table*}

\begin{table*}
    \centering
    \caption{Hardness ratios corresponding to the count rates listed in Table \ref{rates_tab}. The hardness ratios are defined as (1.2--7.0 keV)/(0.3--1.2 keV). }
    \begin{tabular*}{\textwidth}{l @{\extracolsep{\fill}} rrrrr}    
    \hline\hline
        Obs ID & Knot 1 & Knot 2 & Knot 3 & Knot 4 & Combined\\
        \hline
        12238 & $0.78\pm0.33$ & $0.28\pm0.08$ & $\leq0.17$ & $0.57\pm0.49$ & $0.32\pm0.07$ \\
        16028 & $1.27\pm0.50$ & $0.23\pm0.07$ & $0.22\pm0.16$& $\leq0.41$ & $0.35\pm0.07$ \\
        16029 & $0.38\pm0.16$ & $0.27\pm0.09$ & $0.50\pm0.33$& $1.1\pm0.8$ & $0.36\pm0.08$ \\
        Total & $0.70\pm0.15$ & $0.23\pm0.04$ & $0.19\pm0.08$ & $0.28\pm0.18$ & $0.31\pm0.04$ \\
        \hline
    \end{tabular*}
    \label{hr_tab}
\end{table*}

\subsubsection{X-ray core} \label{sc:xray_core}

If, as we suggest, S\,10 is a powerful microquasar within NGC\,300, then we expect to find a hard X-ray source at its core, a signature of accretion. However, we do not observe this; all the knots are best-fitted with soft thermal plasma models, with the majority of emission being $<2\,\kev$. We do find a marginal detection of $6^{+8}_{-5}$ net counts\footnote{Within the 1\barcs5 aperture, we expect $\approx$0.6 background counts, which makes the X-ray emission in the hard band significant at the $\geq99\%$ level, according to Poisson statistics in the presence of a background \citep{1991ApJ...374..344K}.} in the 2--7 keV band, near the central position of the brightest knot (knot 2). However, this is also consistent with the $\approx$5.8 net counts expected from a thermal plasma at a temperature of $0.53\,\kev$, fitted to the spectrum of knot 2. From that, we estimate a 90\% upper limit of $\approx$6 $\times 10^{-5}$ net ct s$^{-1}$ from a point-like power-law core (in addition to the soft thermal emission). For an assumed photon index $\Gamma = 1.7$, this corresponds to a luminosity $L_{0.3-8} \approx 10^{36}$ erg s$^{-1}$. No other knots have significant 2--7$\,\kev$ emission, and thus, regardless of where the central source is, we can take this value as the upper limit of the current accretion luminosity of the central engine.

\begin{figure}
\centering
\includegraphics[width=0.48\textwidth]{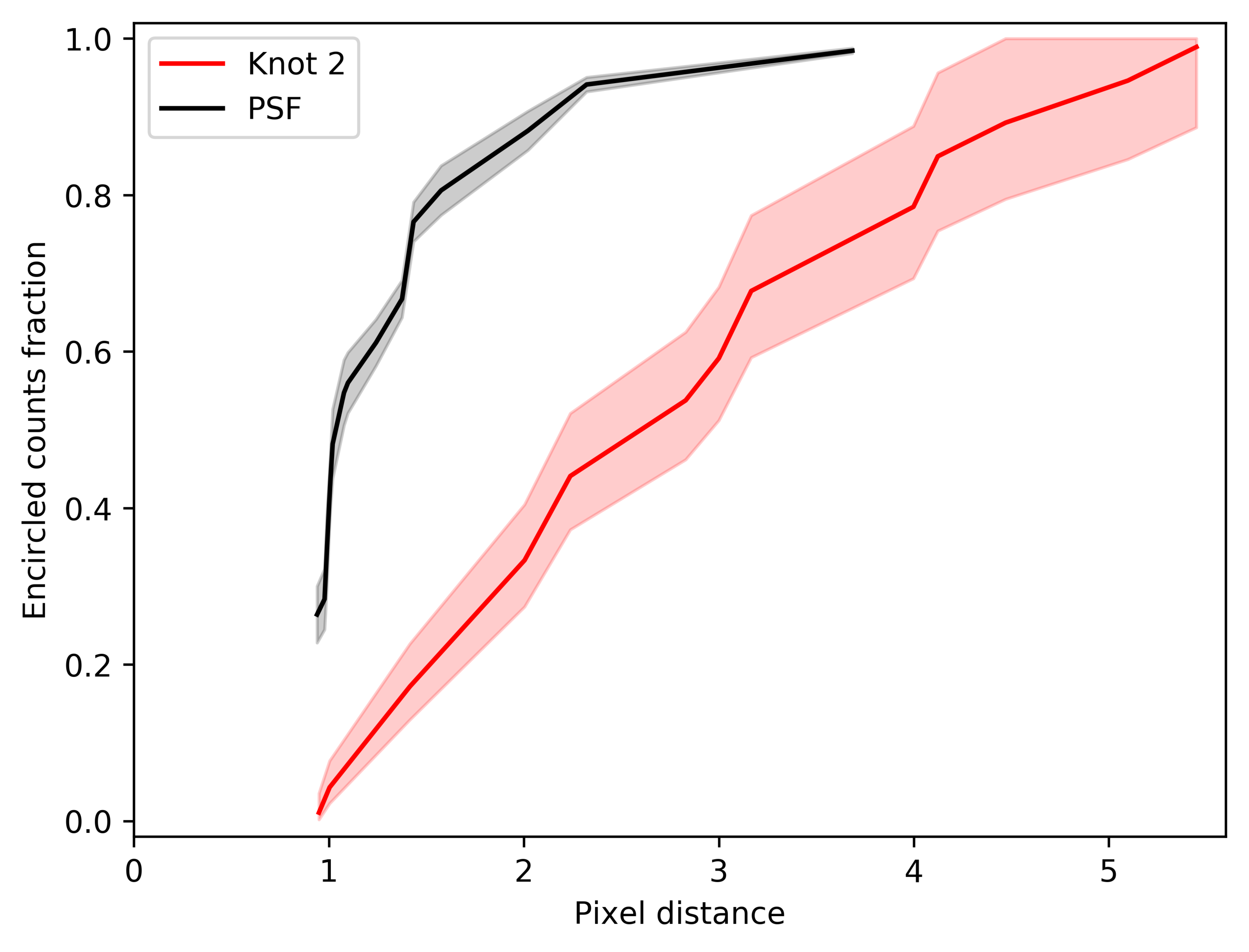}
\vspace{-0.4cm}
 \caption{Enclosed ACIS-I count fraction as a function of distance from the centre of a source. The black points represent the PSF of {\it Chandra}/ACIS-I, from a 1000-iteration Monte Carlo simulation, at the location of S\,10, for ObsID 16028. The red points represent the measured extent of knot 2 from the same observation. Shaded regions indicate 1-sigma errors. At the distance of 1.88 Mpc, 1 ACIS-I pixel corresponds to $\approx$4.5 pc. The comparison shows that the source is significantly extended.}
  \label{psf_im}
\end{figure}

\begin{figure}
\centering
\includegraphics[width=0.32\textwidth,angle=270]{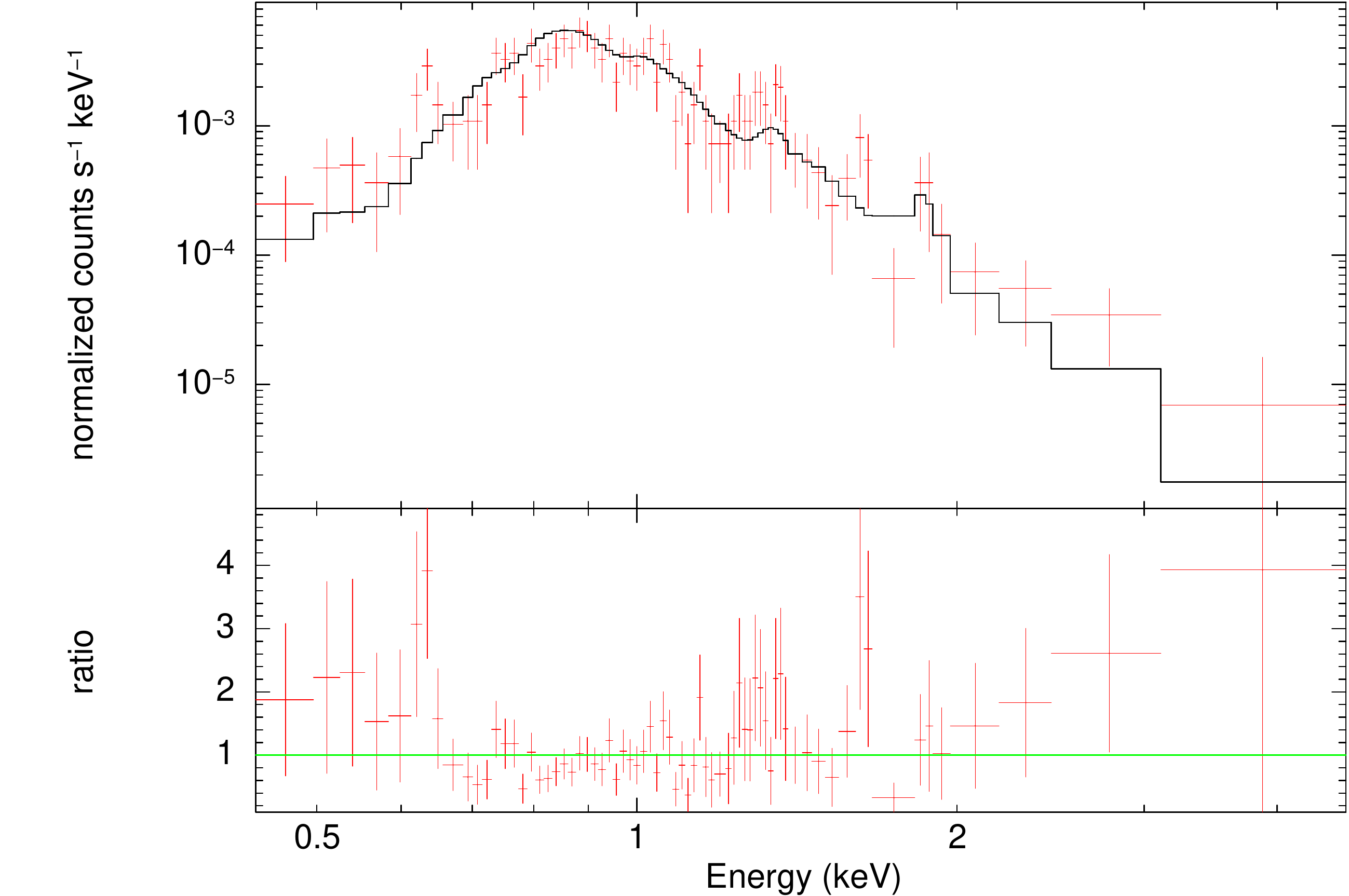}
 \caption{{\it Chandra}/ACIS-I stacked spectrum of the combined knots emission from observations 12238, 16028 and 16029, fitted with a thermal plasma model; the best-fitting plasma temperature is $kT = (0.6 \pm 0.1)\,\kev$. See Table \ref{xray_tab} for the fit parameters.} 
  \label{xray_spectra}
\end{figure}

In the absence of a well-resolved, point-like hard source, we can use indirect arguments to constrain the possible location of the candidate microquasar core. Spatially resolved X-ray jets comprising a series of bright knots are seen in some AGN: most famously, in M\,87 \citep{2002ApJ...568..133W,2002ApJ...564..683M} and NGC\,5128 \citep{2002ApJ...569...54K,2003ApJ...593..169H}. In those cases, the active nucleus is generally located at one end of the string of X-ray knots, because the jet coming towards us is much brighter than the (usually undetected) counter-jet. In the NGC\,300 candidate microquasar, this would correspond to knot 1 (which is also the one with the hardest X-ray colours and hottest mekal fit temperatures). However, the lobe-like morphology of the inner knot of H$\alpha$ emission (pointing towards the south-east: Figure~\ref{opt_im}), and the shape of the larger H$\alpha$ bubble (Figure~\ref{vlt_im}) suggest that knot 1 is unlikely to be the origin of the jet, and that the core is most likely associated with knot 2 (the most luminous part of the system in X-rays, radio and H$\alpha$). We have already shown that knot 2 is extended (Figure \ref{psf_im}), and suggest that it likely consists of two sources separated by $\approx$1.2$\arcsec$ and also aligned in the general direction of the other knots. Thus, we suggest that knot 2 consists of a pair of ejecta or hot spots from the approaching and receding jet and that the core resides somewhere between the two. 

\begin{table*}
    \centering
    \caption{Best-fitting parameters for the spectra of the individual knots and for the combined spectrum, from the merged {\it Chandra}/ACIS-I observations 12238, 16028 and 16029, fitted with {\it TBabs} $\times$ {\it mekal}. We fixed the absorbing column density $N_{\rm H}$ to the Galactic line-of-sight value; adding intrinsic local absorption does not produce statistically significant improvements. Errors are 90\% confidence limits for one interesting parameter. We report the observed fluxes and de-absorbed luminosities.} 
    \begin{tabular*}{\textwidth}{l @{\extracolsep{\fill}} lrrrrr}
        \hline\hline
        Model & Parameter & Knot 1 & Knot 2 & Knot 3 & Knot 4 & Combined\\
        \hline
        & Net Counts (0.3--7.0 keV) & 99 & 234 & 58 & 26 & 432\\\hline
        TBabs & $N_{\rm H}\Big(10^{22}\,\centi\meter^{-2}\Big)$ & [0.03] & [0.03] & [0.03] & [0.03] & [0.03] \\[2ex]
        Mekal & $k_eT$ (\kev) & \U{0.80}{0.15}{0.14} & \U{0.53}{0.07}{0.10} & \U{0.37}{0.15}{0.08} & \U{0.28}{0.31}{0.07} & \U{0.57}{0.05}{0.06} \\[2ex]
         & K $\Big(10^{-6}\Big)$ & \U{2.1}{0.4}{0.3} & \U{5.8}{1.3}{0.8} & \U{1.8}{0.9}{0.6} & \U{1.2}{1.3}{0.7} & \U{9.2}{1.0}{0.9}\\[2ex]
        & $f_{0.3-8.0}$ $\Big(10^{-15}\flux\Big)$ & \U{4.7}{0.8}{0.8} & \U{14.4}{1.9}{1.9} & \U{3.8}{1.1}{1.1} & \U{2.1}{1.3}{1.3} & \U{22.9}{2.2}{2.2}\\[2ex]         
        & $L_{0.3-8.0}$ $\Big(10^{36}\,\ergs\Big)$ & \U{2.2}{0.4}{0.4} & \U{6.9}{0.9}{0.9} & \U{1.9}{0.6}{0.6}& \U{1.1}{0.7}{0.7} & \U{11.0}{1.0}{1.1} \\[2ex]
        & w-statistic/d.o.f. & 79.3/55 & 64.2/73 & 37.6/37 & 14.1/15 & 115.8/108\\
        \hline
    \end{tabular*}
    \label{xray_tab}
\end{table*}



\subsection{Radio results} \label{sc:radio_results}

From the ATCA images, we discovered a bright, resolved and elongated radio structure aligned along the axis of the X-ray knots (Figure \ref{xray_im}). The majority of the radio emission comes from a region coincident with the first and second X-ray knots, with the radio peak coincident with X-ray knot 2. In the 5.5\,GHz image, we also detect a region of enhanced radio emission, which we refer to as a `radio node', north-west of the bright, central radio component, near X-ray knot 4; the same radio node is a marginal detection at 9 GHz. From the 5.5\,GHz image, we measure a total size of the elongated radio bubble as  $19'' \times 6'' \approx 170 \times 55\,\parsec$ (for a distance of 1.88 Mpc). This emitting region is clearly resolved both along and across the jet direction, because the beam size is only $\approx$26 $\times$ 13\,pc (Section \ref{rad_dat_red_sec}). 

We find total integrated flux densities of $f_{5.5\,\GHz}=520\pm15\,\micro\jansky$ and $f_{9\,\GHz}=380\pm20\,\micro\jansky$ for the 5.5 and 9\,GHz bands, respectively. This equates to a 5.5\,GHz radio luminosity of $L_{\rm 5.5\,\GHz}\approx1.2\times10^{34}\,\ergs$. We fitted a 2-dimensional Gaussian profile to the central, and brightest, region of S\,10 using the CASA task {\sc imfit}, to estimate the flux density coming from the core. We find peak intensities of $87.8\pm2.7\,\ujy$\perbeam and $35.0\pm2.8\,\ujy$\perbeam at 5.5 and $9\,\GHz$ respectively, and integrated flux densities of $429\pm30\,\ujy$ and $371\pm29\,\ujy$. We also used a 2-dimensional Gaussian profile fit in the 5.5\,GHz map to measure the integrated flux of the radio node to the northwest of the core; we find a flux density of $79.5\pm5.3\,\ujy$. The knot is only marginally detected in the 9\,GHz image, with an integrated flux density of $17.9\pm4.4\ujy$. Our measured radio fluxes are summarised in Table \ref{radio_tab}.

The spectral index $\alpha$ (defined as $f_{\nu} \propto \nu^{\alpha}$)  ranges from about $-0.7$ to about $-0.4$ near the central region (where the radio emission is strongest), at the location of X-ray knot 2 (Figure \ref{radio_im}). This is consistent with optically-thin synchrotron emission. There are also hints of a flattening of the spectral index towards the edges (near X-ray knots 1 and 3). We also measured the spectral index of the radio node at the north-west end of the bubble, finding $\alpha = -1.3\pm0.9$ there\footnote{The knot is only marginally detected at $9\,\GHz$. To determine its spectral index, we used the region defined by the 5.5\,GHz {\sc imfit} output and integrated the flux over this same area for the 9\,GHz image.}. It is unclear whether this radio node is associated with the X-ray knot 4, as the peak radio emission (in the 5.5\,GHz band) of the radio node appears to be slightly offset to the north-west of the X-ray knot (Figure \ref{xray_im}, top panel).

We also detect radio emission from the nearby, possibly unassociated H{\sc ii} region H\,10 \citep{1997ApJS..108..261B}, south-west of S\,10 (Figure \ref{radio_im}). The spectral index in this region is much flatter than in the candidate microquasar bubble (Figure \ref{radio_im}), as expected for the spectral index of an H{\sc ii} region. At $5.5\,\GHz$, the peak flux density of the H{\sc ii} region is $(52.6\pm2.7)\,\ujy$\perbeam and the integrated flux density is $(511\pm15)\,\ujy$. At $9\,\GHz$, 
the peak flux density is $(30.1\pm2.8)\,\ujy$\perbeam and the integrated flux density is $(302\pm24)\,\ujy$. A few arcsec further to the south-west, another region with steep spectral index corresponds to the supernova remnant catalogued as S\,9 in \cite{1997ApJS..108..261B}. A study of H\,10 and S\,9 is beyond the scope of this work.

\subsection{Optical results} \label{sc:optical_results}

\subsubsection{Diffuse emission} \label{sc:diffuse_emission}

\citet{1997ApJS..108..261B} used CCD imaging and spectroscopic data taken in 1987 with the 2.5-m du Pont telescope to measure H$\alpha$ and [S {\sc ii}] fluxes of optical SNR candidates in NGC\,300. Their source S\,10 = DDB2 corresponds to the X-ray and radio structures discussed in this paper. S\,10 is listed in the catalogue of \citet{1997ApJS..108..261B} with a line ratio [S {\sc ii}]:(H$\alpha$ plus [N {\sc ii}]) $\approx 0.67$, indicative of shock-ionized gas, and with an H$\alpha$ flux $F_{\rm H\alpha} \approx 8.3\E{-15}\,\flux$ \citep[][from the average surface brightness and size, in their Table 3A]{1997ApJS..108..261B}. This value has already been corrected for spectral contamination from [N {\sc ii}]$\lambda \lambda$6548,6583, which \citet{1997ApJS..108..261B} assumed to be 25\% of the H$\alpha$ emission, based on an average value over their sample of SNR candidates in NGC\,300. In fact, individual long-slit spectra of S\,10 also from \citet{1997ApJS..108..261B} show a slightly larger [N {\sc ii}] contribution of $\sim$35\%, implying an H$\alpha$ flux $F_{\rm H\alpha} \approx 7.7\E{-15}\,\ergs$.
The diameter of this line-emitting region was reported by \citet{1997ApJS..108..261B} as only $\approx$14 pc\footnote{Rescaled to our adopted distance of 1.88 Mpc; \citet{1997ApJS..108..261B} used a value of 2.1 Mpc.}. This is much smaller than the size of the H$\alpha$ structure we see for example in the VLT images (Figure~\ref{vlt_im}). We suspect that the S\,10 measurements of \citet{1997ApJS..108..261B} refer only to the brightest core of the nebula, near X-ray knot 2, not to the larger structure around it. Thus, we expect the total H$\alpha$ flux from the whole region to be a few times larger.  

In 2006, follow-up spectroscopic observations of NGC\,300 by \citet{2011Ap&SS.332..221M}, with the 2.3-m Advanced Technology Telescope at Siding Spring Observatory, indicated a slightly lower [S{\sc ii}]:H$\alpha$ ratio of $0.35\pm0.15$ for S\,10, although the source was still classified as an SNR based on its multiband properties. More importantly, \citet{2011Ap&SS.332..221M} found that the S\,10 emission region is significantly larger (diameter of $\approx$56 pc) than originally reported, and an order of magnitude more luminous ($F_{\rm H\alpha} = (10.5\pm1.4)\E{-14}\,\flux$). This flux value was based on the extrapolation to the whole nebula of the integrated line flux measured from a long-slit observation; such extrapolations are fraught with uncertainties. Moreover, at the average seeing of Siding Spring Observatory, S\,10 is inevitably contaminated by emission from the nearby H\,10 H{\sc ii} region, which may explain the lower value of [S{\sc ii}]:H$\alpha$ measured by \citet{2011Ap&SS.332..221M}. Size and flux discrepancies for several other NGC\,300 sources between \citet{1997ApJS..108..261B} and \citet{2011Ap&SS.332..221M} are also discussed in the latter paper.

To improve on those previous two studies and resolve their discrepancies, we investigated the diffuse optical emission at S\,10 using archival {\it HST}/ACS images in the F475W, F606W and F814W filters (Figure \ref{opt_im}). This is possible because the broadband filter F606W also includes the wavelengths around H$\alpha$; therefore, we expect regions of strong H$\alpha$ emission to show up in green, in a true-colour image from those three bands. There are indeed two bright regions of H$\alpha$ emission (Figure \ref{opt_im}, bottom panel), separated by a distance of $\approx$20 pc, on either side of the central location of X-ray knot 2. 
The southern H$\alpha$ source has a lobe-like structure, which suggests that the gas has been shocked by some kind of fast ejection: either a microquasar jet or collimated ejecta during a SN explosion. The lobe-like source has a diameter of $\approx$0\barcs8, corresponding to $\approx$15\,pc; it is aligned along the same direction defined by the string of X-ray knots and by the elongated radio source. This is most likely the source identified and measured by \citet[][their Fig.~3]{1997ApJS..108..261B}.

To investigate the presence of fainter, much more extended H$\alpha$ emission, and to measure its flux, we used the continuum-subtracted VLT image. Although at lower spatial resolution, we clearly recover (Figure \ref{vlt_im}) the two innermost H$\alpha$ emission regions already seen in the broadband {\it HST} image. Additionally, we see an extended bubble, stretching between and around X-ray knots 1 and 2 (characteristic size of $\approx$60 pc), with a brighter spot almost coincident with X-ray knot 1 (Figure \ref{vlt_im}). The emission region is bounded on its southern side by a slightly brighter rim. We speculate that this H$\alpha$-emitting bubble has been inflated by a source of kinetic power probably located near X-ray knot 2: the same source of kinetic power responsible for the radio and X-ray emission.

To measure the flux of the extended H$\alpha$ region, we defined suitable source and background regions in the continuum-subtracted VLT image (Figure \ref{vlt_im}), with the imaging and photometry tool {\sc DS9}. We then used the VLT FORS2 Exposure Time Calculator in Imaging Mode\footnote{https://www.eso.org/observing/etc/} to convert from the measured net count rate to a physical line emission flux. We assumed a contribution of [N {\sc ii}]$\lambda \lambda$6548,6583 equal to $(30\pm5)$\% of the H$\alpha$ flux. This choice is based on the line ratios measured directly for S\,10 by \citet{1997ApJS..108..261B} and \citet{2011Ap&SS.332..221M}, and takes also into account the small fraction of [N {\sc ii}] flux that falls outside the VLT filter passband. We obtain a flux $F_{\rm H\alpha} = (4.1\pm0.5)\E{-14}\,\flux$. The error is the quadrature sum of the uncertainty on the [N {\sc ii}]$\lambda \lambda$6548,6583 contribution and a rough estimate of the range of values we obtained for slightly different choices of the outer boundary of the S\,10 bubble. We also verified this flux estimate by estimating the ratio (a factor of 5) between the H$\alpha$ counts in the whole bubble and in the bright core, for which \citet{1997ApJS..108..261B} had reported a flux of $\approx$8$\E{-15}\,\flux$. Finally, we confirmed the consistency of our estimate by comparing our measured VLT count rates for S\,10 and for isolated SNR candidates that have a more reliable flux measurement in  \citet{1997ApJS..108..261B}; we obtained the same estimate of an S\,10 H$\alpha$ flux between $\approx$4--5$\E{-14}\,\flux$.  At the distance of 1.88 Mpc, this corresponds to a luminosity $L_{H\alpha}=(1.7\pm0.2)\E{37}\,\ergs$ and (for a standard Balmer decrement from Case B recombination) $L_{H\beta}=(6\pm1)\E{36}\,\ergs$.


\begin{table}
    \centering
    \caption{Integrated ATCA 5.5 amd 9\,GHz fluxes. For fluxes values obtained via the CASA task {\sc imfit} we provide spectral indices. The steep spectral index of the core suggests that the emission is coming from optically-thin synchrotron emission, as expected from other radio bubbles.}
    \begin{tabular*}{0.48\textwidth}{l @{\extracolsep{\fill}} rrr}
        \hline\hline
        \multicolumn{1}{c}{Object} & \multicolumn{1}{c}{5GHz flux}& \multicolumn{1}{c}{9GHz flux} & \multicolumn{1}{c}{$\alpha$}\\
         & \multicolumn{1}{c}{$\micro\jansky$} & \multicolumn{1}{c}{$\micro\jansky$} & \\
        \hline
        Total & $520\pm15$ & $380\pm20$ &  \\
        Core & $429\pm30$ & $352\pm29$ & $-0.4\pm0.2$\\
        Node & $75\pm7$ & $39\pm17^*$ & $-1.3\pm0.9$ \\
        H {\sc ii} Region & $511\pm15$ & $302\pm24$ & \\
        \hline
    \end{tabular*}
    \label{radio_tab}
    \begin{flushleft}
    $^*$ Flux integrated over region defined by radio node in $5.5\,\GHz$ image\\
    \end{flushleft}
\end{table}

\subsubsection{Stellar counterparts} \label{sc:stellar_count}

We selected a few potential optical counterparts in the {\it HST} images (Figure \ref{opt_im}) located close to knot 2, as this is the most likely location of the true nuclear source (as discussed in Section \ref{sc:xray_core} and from the H$\alpha$ morphology discussed in Section \ref{sc:diffuse_emission}); we measured their optical brightnesses and colours with {\sc iraf}. We interpreted their physical properties by comparing those values with the Padova theoretical stellar isochrones\footnote{Available at http://stev.oapd.inaf.it/cgi-bin/cmd} \citep{2012MNRAS.427..127B,2015MNRAS.452.1068C} for a metallicity $Z = 0.015$ (Table \ref{stars_tab}). Specifically, we estimated the age, mass, radius and temperature on the nearest isochrone to each star. In doing so, we assumed that any emission we see is due to a single star and not to an accretion disc around the compact object or to multiple stars.  Star A is the bluest and brightest stellar source in the field, a B-type giant; however, it appears unlikely to be the true optical counterpart of the central engine, as it resides outside of a lobe-shaped H$\alpha$ nebula (Figure \ref{opt_im}), and therefore, offset from the likely direction of the jet that inflated that structure\footnote{If we attribute the offset from the jet axis to proper motion of the star after the main jet activity phase, we require a projected velocity of $\approx$100 km s$^{-1}$ for $\approx$10$^5$ yr.}. Star C is also outstanding in brightness and colour from the rest of the stellar population in the field: it is consistent with an intermediate-age AGB star (near the top of the AGB branch), and it lies along the direction of the jet. Perhaps the most intriguing candidate optical counterpart is star D. It is also located along the direction of the jet, approximately half way between the two peaks of H$\alpha$ emission (Figure \ref{opt_im}). It is detected as a moderately bright source only in the F814W image (Vegamag $m_{814} \approx 23.1$ mag). Its extremely red colour ($m_{606}-m_{814} > 2.6$ mag;$ m_{475} - m_{814} > 3.2$ mag) suggest a high intrinsic reddening, higher than for any other neighbouring star; this could be due for example to circumstellar dust. For example, among the possible alternative interpretations, we cannot rule out that star D is a hypergiant or Wolf-Rayet star with intrinsic absolute brightness $M_V < -7$ mag, but a strong extinction $A_V \gtrsim 7$ mag. We shall mention later (Section 4.2) that a massive donor star surrounded by shells of ejected material is consistent with one of the scenarios discussed for the feeding of this candidate microquasar. Further investigations on the nature of star D is beyond the scope of this paper. A discussion on the relative contributions of accretion disk and donor star to the optical emission is also left to further work.

We also searched for unusually bright, point-like optical sources at the location of the other X-ray knots, but did not find any unusual candidate. Most of the other stars visible in the {\it HST} image of the field (Figure \ref{opt_im}) are red giants from an intermediate-age population (age of $\sim$ a few $10^8$ yr), with a few B-type main sequence stars from a younger population.

\begin{table*}
    \centering
    \caption{{\it HST} magnitudes and physical parameters of four potential optical counterparts of the true nuclear source of S\,10. The optical counterparts (along with the lettering) corresponds to those outlined in Figure \ref{opt_im}. Observed magnitudes are converted to physical properties using the theoretical isochrones for a metallicity $Z = 0.015$. The stellar radii are calculated using the surface gravity and total mass. Colours for star D are dominated by intrinsic reddening and thus comparison to theoretical isochrones is meaningless.}
    \begin{tabular*}{0.96\textwidth}{l @{\extracolsep{\fill}} rrrrrrrr}
        \hline\hline
        \multicolumn{1}{c}{Star} & \multicolumn{1}{c}{F814W} &\multicolumn{1}{c}{F606W} & \multicolumn{1}{c}{F475W} & \multicolumn{1}{c}{Age}& \multicolumn{1}{c}{Mass}& \multicolumn{1}{c}{Radius} & \multicolumn{1}{c}{Temperature} & \multicolumn{1}{c}{Luminosity} \\
         & \multicolumn{1}{c}{mag}& \multicolumn{1}{c}{mag}&\multicolumn{1}{c}{mag}& \multicolumn{1}{c}{Myr} & \multicolumn{1}{c}{$\Msol$} & \multicolumn{1}{c}{$\Rsol$} & \multicolumn{1}{c}{$\Kelvin$} & \multicolumn{1}{c}{$\ergs$} \\
        \hline
        A & $-3.67\pm0.02$ & $-3.86\pm0.02$ & $-3.88\pm0.02$ & $\sim50$ & 7 & 110 & 12,300 & $2\E{37}$  \\
        B & $-2.77\pm0.07$ & $-2.10\pm0.05$ & $-1.86\pm0.06$ &$\sim200$ & 4 & 120 & 6,600 & $2\E{36}$ \\
        C & $-4.91\pm0.01$ & $-2.30\pm0.06$ & $-0.66\pm0.05$ &$\sim500$ & 3 & 1,600 & 3,400 & $2\E{37}$ \\
        D & $-3.30\pm0.04$ & $\geq-0.7$ & $\geq-0.1$ & N/A & N/A & N/A & N/A & N/A\\ 
        \hline
    \end{tabular*}
    \label{stars_tab}
\end{table*}

As in our ATCA data, the H{\sc ii} region to the south-west, H\,10 \citep{1997ApJS..108..261B}, is clearly visible. The {\it HST} images show that there is strong F606W emission coincident with a population of bright, blue stars (Figure \ref{opt_im}).


\section{Discussion} \label{sc:discussion}


We have identified a chain of four thermal X-ray sources aligned with each other and with an elongated 170-pc-long radio structure. The X-ray and optical sources are associated with an optical nebula originally identified as an SNR, but which is also reminiscent of jet-powered ULX bubbles. Here, we argue that the most likely explanation for the multiband structure is that S\,10 is a powerful microquasar, although we also consider alternative explanations. The X-ray spectra for the knots are dominated by optically-thin thermal plasma emission, likely a result of jet/ISM interactions. The radio emission has a steep spectral index, consistent with optically-thin synchrotron emission; this is consistent with a relativistic electron population energized by a microquasar jet. Finally, the associated optical line-emission regions are shock-ionised, consistent with a forward shock propagating into the ISM, caused by the impact of a microquasar jet or fast SN ejecta. This is displayed in Figure \ref{jet_im}, where the {\it Chandra}, ATCA and VLT images have been aligned and rotated to demonstrate the connections between structures observed in each band. In this section, we collate and examine in more detail the observational properties of S\,10 that have led to our microquasar interpretation. Finally, we attempt to determine the jet power of S\,10.

\subsection{Emission processes in the knots}


An important finding of our {\it Chandra} analysis is that the X-ray emission is from thermal plasma. This is likely the result of jet-driven shocks propagating into the ISM. The density inferred for the X-ray emitting hot plasma (Section \ref{xray_results_sec}) is consistent with typical pre-shock ISM densities $\sim$ 1 cm$^{-3}$. It is hard to compare the knotty structure of the NGC\,300 candidate microquasar (Figure \ref{jet_im}) with the fine details of X-ray jet knots seen in nearby AGN such as M\,87 and NGC\,5128, because of the much higher spatial resolution (relative to jet length) and signal-to-noise ratio for the latter class of jets. Instead, the {\it Chandra} images of the S\,10 jet are reminiscent of the X-ray appearance of jets in higher-redshift radio galaxies and quasars \citep[e.g.,\,][]{2011ApJS..197...24M,2015ApJS..220....5M,2016ApJ...833..123M}. However, appearances can be deceptive. The X-ray emission from AGN and quasar jets (including the emission from their hot spots) is almost always non-thermal \citep{2006ARA&A..44..463H}: either synchrotron radiation from relativistic electrons (with Lorentz factors $\gamma \gtrsim 10^7$), or inverse Compton scattering of cosmic microwave background photons off slightly less energetic electrons ($\gamma \sim 10^3$). The only few exceptions, where X-ray hot spots have been attributed to thermal emission, are sources in which the AGN jet is colliding with a dense, cold cloud; {\it e.g.}, the jets in the radio galaxies PKS\,2152$-$699 \citep{2005ApJ...618..609L}, 4C$+$29.30 \citep{2012ApJ...750..124S}, and 3C 277.3 \citep{2016MNRAS.458..174W}. Instead, in both of the known off-nuclear microquasars (NGC\,7793 S\,26 and NGC\,300 S\,10) with evidence of X-ray jets, the hot spots are dominated by thermal emission at temperatures $\sim$0.5 keV. In both sources, the total X-ray emission from shocked gas is $\sim2\E{37}\,\ergs$. Another iconic microquasar, the Galactic source SS\,433, also has a hot spot consistent with thermal emission, along the eastern jet \citep{2007A&A...463..611B} (instead, the termination shock further downstream is non-thermal). We do not have enough sources to determine whether the higher contribution of thermal emission in microquasar hot spots is due to a comparatively higher ISM density, or to a different composition and Lorentz factor in the jet.


Another key observational property of S\,10 is the elongated radio structure aligned with the chain of X-ray knots. How the radio and X-ray emission are linked is unclear: they arise from differing emission mechanisms (synchrotron and thermal-plasma respectively) but share similar morphologies. In the 5.5\,GHz ATCA image, we detect a radio node beyond (further downstream along the jet) the outer X-ray knot (Figure \ref{xray_im}). \citet{2010MNRAS.409..541S} find that the X-ray hotspots of S\,26 slightly ($\approx20\,\parsec$) trail the radio lobes, further evidence that they arise from different physical processes. We cannot determine whether the outer X-ray knot and the radio node of S\,10 are related (Figure \ref{jet_im}): the true X-ray counterpart to the radio node (if there is one) may have already faded. 



 

\subsection{Origin of the discrete knot structure}


\begin{figure}
\centering
\includegraphics[width=0.465\textwidth]{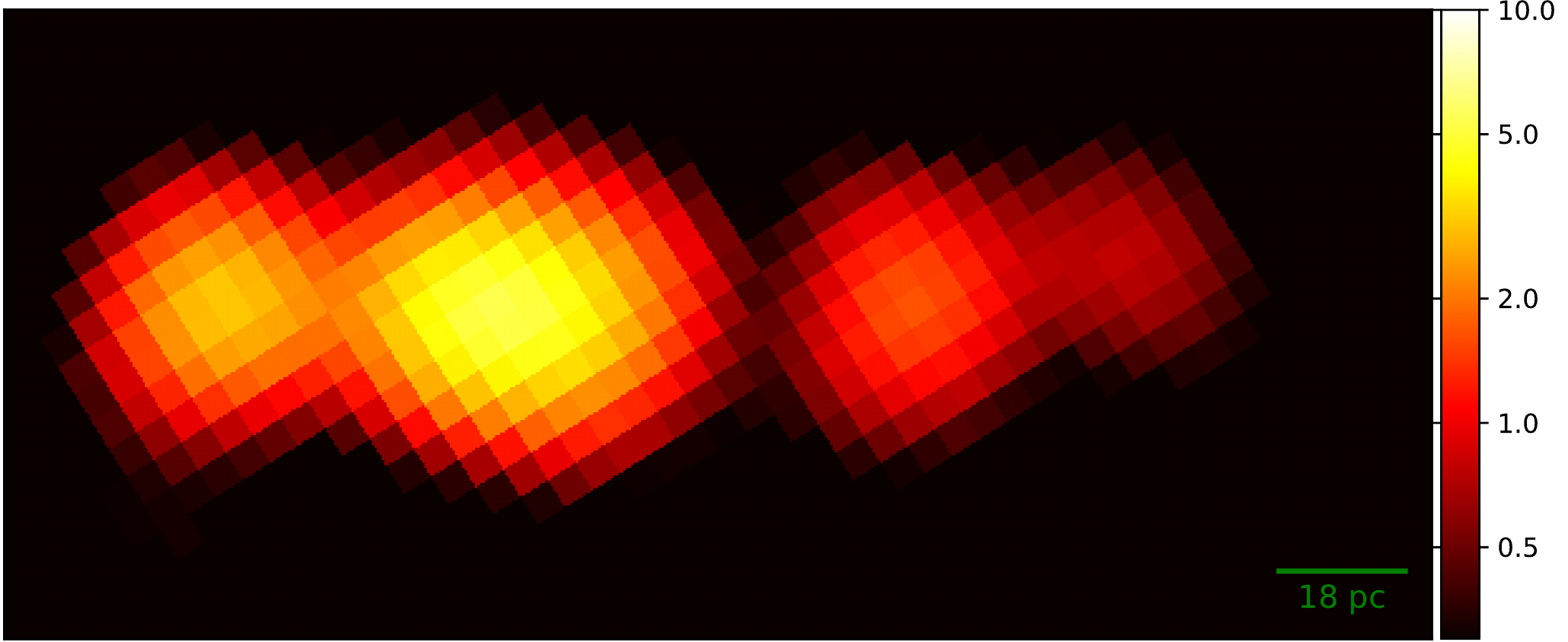}\\
\includegraphics[width=0.465\textwidth]{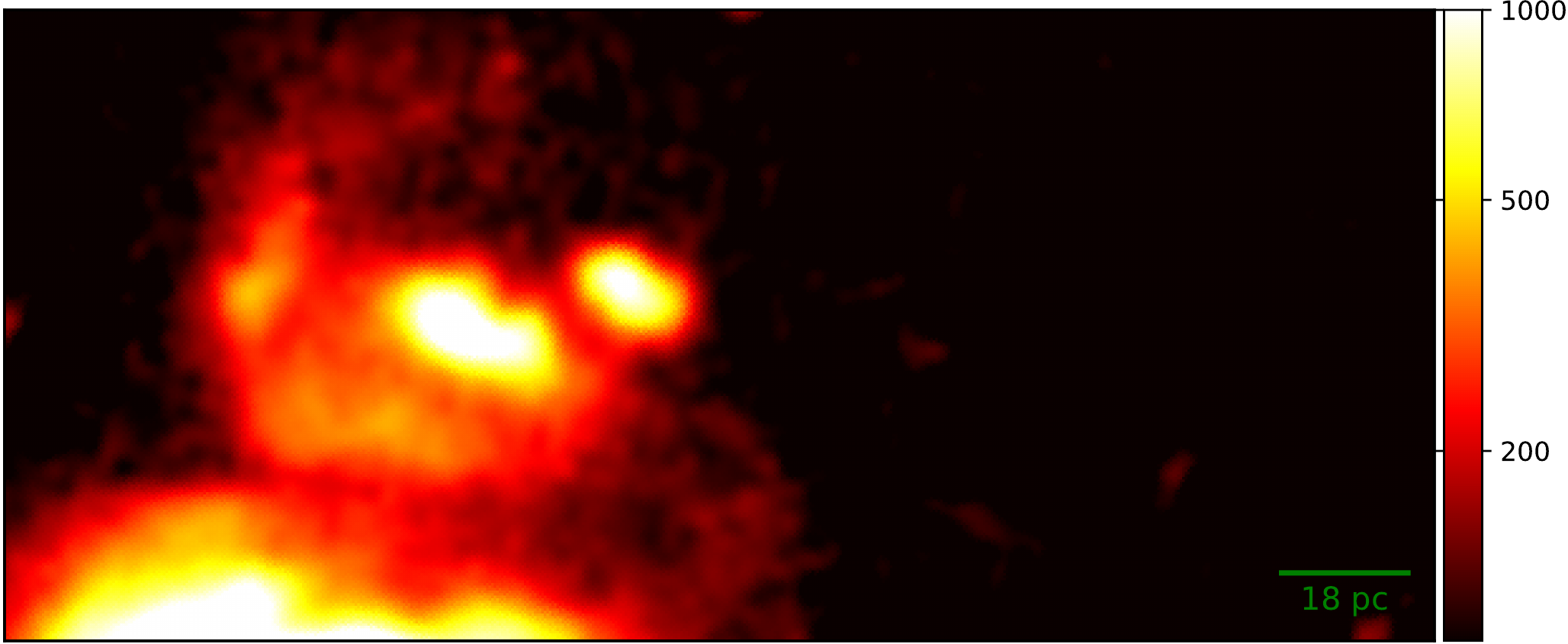}\\
\includegraphics[width=0.465\textwidth]{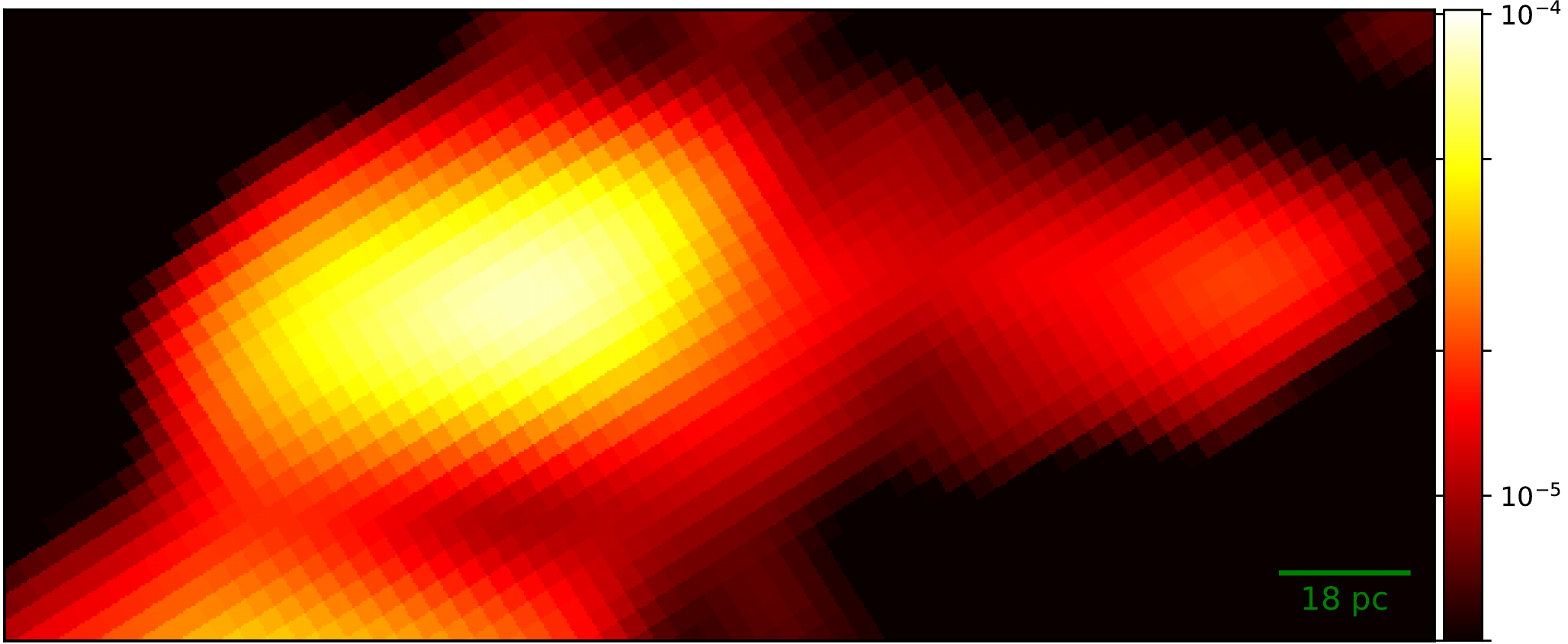}\\
\hspace{-0.15cm}
\includegraphics[width=0.472\textwidth]{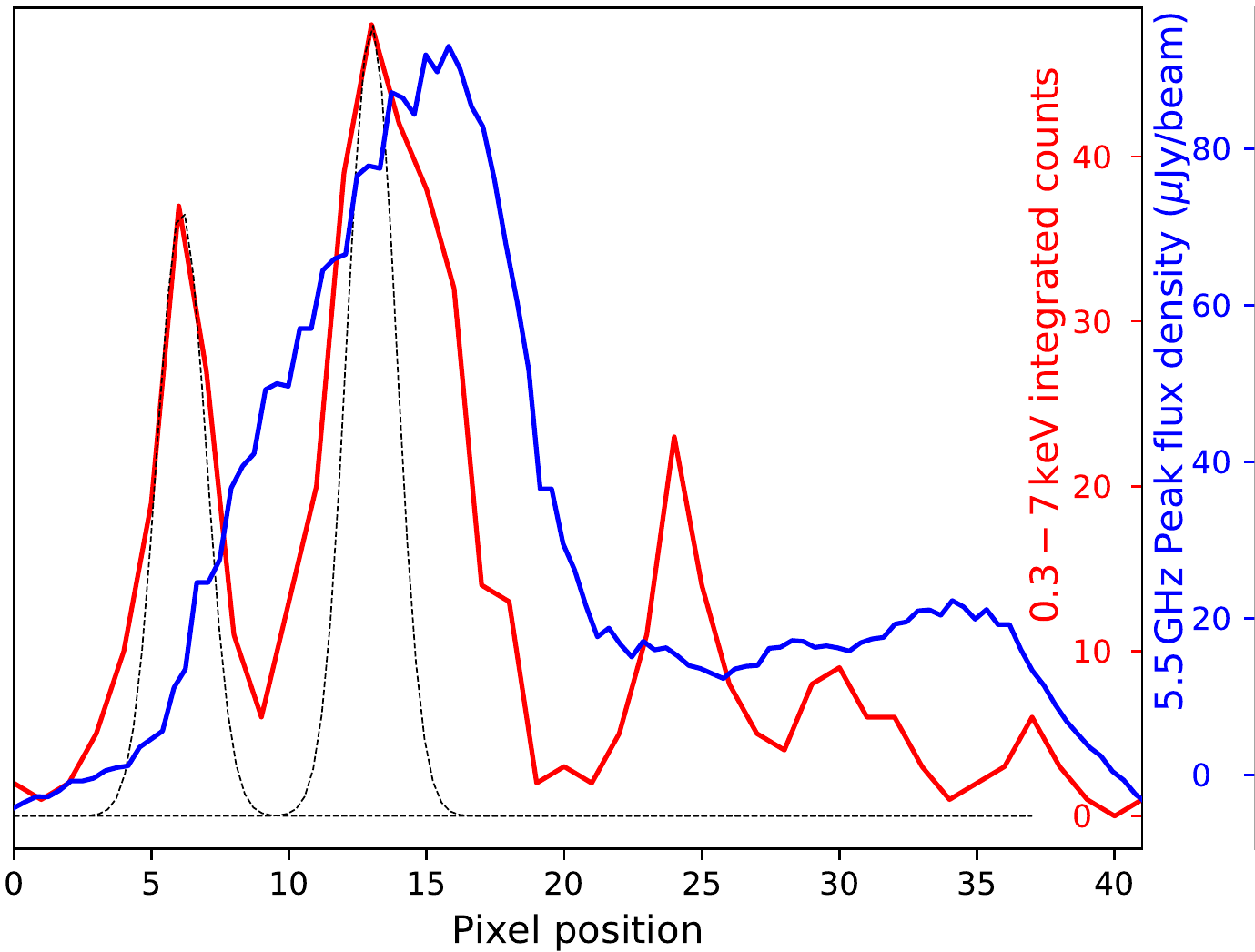}\\
 \caption{From top to bottom: S\,10 jet in X-rays ($0.3-7\,\kev$; scalebar in units of counts), H$\alpha$ line (scalebar in units of ADU for the VLT/FORS2 CCD), and 5.5\,GHz band (scalebar in units of Jy\,beam$^{-1}$), rotated to a horizontal position; and corresponding brightness distribution along the jet in X-rays (red curve) and radio (blue curve), on the same spatial scale. For the bottom panel, we measured the X-ray brightness by integrating the pixel counts in 1-pixel-wide strips perpendicular to the jet axis; we measured the radio intensity from the peak value along the jet. The dotted black line represents a symmetric Gaussian approximation of the {\it Chandra} PSF of a point-like source at the location of S\,10, determined from our PSF modelling (Section \ref{xray_results_sec}). At the distance of 1.88 Mpc, 1 ACIS-I pixel corresponds to $\approx$4.5 pc. The comparison shows that the second X-ray knot is extended and likely composed of (at least) two distinct sources, displaced by $\approx$1\barcs2 along the direction of the jet. For comparison, we show that the first knot instead matches the {\it Chandra} PSF.}
  \label{jet_im}
\end{figure}

The presence of multiple thermal X-ray knots rather than only two termination hot spots is the main difference between NGC\,300 S\,10 (Figure \ref{jet_im}) and NGC\,7793 S\,26 \citep{2010MNRAS.409..541S}. In this section, we outline possible scenarios that can give rise to a string of bright knots along the jet.  
\newline\\
\noindent {\it (i) Discrete relativistic ejecta:}\\
Numerous compact accreting sources produce large-scale moving jets in the form of discrete ejecta. Ejection of relativistic plasma knots that emit via synchrotron processes have been observed at radio and X-ray wavelengths in the sub-Eddington low-mass X-ray binaries (LMXBs) H1743-322 \citep{2005ApJ...632..504C}, XTE J1550-564 \citep{2002Sci...298..196C, 2003ApJ...582..945K} and XTE J1752-223 \citep{2012MNRAS.423.2656R}. On the super-Eddington side, the ULX Ho II X-1 shows evidence of multiple discrete radio ejecta \citep{2014MNRAS.439L...1C}. While each of the Ho II X-1 knots are not detected in X-rays, the S\,10 jet could be made up of similar discrete ejecta where each southeast-northwest knot pair is due to a flare or outburst. In this scenario, the most recent ejecta would be the two unresolved blobs that constitute knot 2; the previous outburst would have produced knots 1 and 3; knot 4 would be the oldest ejection event whose southeastern counterpart has since faded (Figure \ref{jet_im}). 
However, we rule out this scenario as the X-ray knots would have to be radiating via synchrotron emission, not thermally as we see for S\,10.
\newline\\
\noindent {\it (ii) Internal shocks:}\\
The multiple-knot morphology may be a result of internal shocks within the jet \citep{1994ApJ...430L..93R,2014MNRAS.443..299M}. Faster-moving ejecta catch up to slower-moving, previously ejected material further downstream, the collision causing shocks and accelerating particles. This would make S\,10 analogous to the AGN jets we see in M87 \citep{1978MNRAS.184P..61R, 1996ApJ...473..254S} and 3C 264 \citep{2015Natur.521..495M}. As with the previous scenario, internal shocks would produce synchrotron emission in the X-ray band and thus we can rule out this interpretation.
\newline\\
\noindent {\it (iii) Multiple layers of ISM:}\\
If instead we are looking at a more steady, continuous jet, with a persistent flow, then the additional knots could be explained by the jet punching through different layers of the ISM. The jet may be passing through shells of denser ISM, shocking the gas as it penetrates each layer. The resulting shocked gas would produce thermal X-ray emission. These over-densities in the ISM may simply be a result of random fluctuations in the ISM, or could be shells of material ejected by the massive progenitor either via stellar winds, or in giant eruptions, or during the SN explosion that produced the compact object. Nebulae around even the most powerful Luminous Blue Variable (LBV) stars are $\lesssim$ a few pc in size \citep{2011BSRSL..80..440W}, an order of magnitude smaller than the size of the jet in S\,10. However, the presence of numerous OB stars and likely SNRs in the nearby H {\sc ii} region H\,10 leaves open the possibility that stellar activity created supershells and filaments in the ISM on scales of $\sim$100 pc.
\newline\\
\noindent {\it (iv) Multiple outbursts:}\\
The accretion rate and/or the kinetic power carried by the jet may not be steady; the system may undergo state transitions. This could also be the reason for the current low X-ray luminosity of the core (further discussed in Section 4.4). If this is the case, then we do not expect a continuous jet, but rather phases of enhanced activity. Recurrent outbursts may be creating pairs of hotspots in a Sedov-Taylor phase that keep expanding and cooling as they propagate out from the central object. In between outbursts, the channel drilled by the jet during the previous phase of activity would be refilled by the ISM. The innermost X-ray hotspots, that is the two marginally resolved sources that form {\it Chandra} knot 2, would be those created during the most recent epoch of jet activity. The uneven number of knots may be due to an over density to the east of S\,10 resulting in a pile-up of multiple knots. In AGN, a similar scenario of intermittent activity has been proposed to explain the so called ``double-double radio galaxies'', characterised by multiple pairs of hot spots (often aligned along the same direction) sharing the same core \citep{2000MNRAS.315..371S,2000MNRAS.315..381K,2009BASI...37...63S,2011MNRAS.410..484B}. In this scenario, NGC\,300 S\,10 would be the first unambiguous example of a double-double microquasar.

The fading and/or appearance of new knots would undoubtedly result in X-ray variability. Two or three major episodes of enhanced activity over a timescale of a few 10$^5$ yr correspond to a recurrence timescale too long to be explained by thermal-viscous accretion disk instabilities, for a stellar-mass compact object. It is instead consistent, for example, with the timescale of thermal pulsations in an AGB donor star \citep[{\it e.g.},][]{2013MNRAS.434..488M}, or of giant eruptions in LBV stars and other types of SN impostors \citep[e.g.,\,][]{2006ApJ...645L..45S,2009ApJ...699.1850B,2015MNRAS.447..117T}. In particular, if the massive star is in a binary system with a neutron star or a black hole, giant eruptions are expected to produce phases of highly super-Eddington mass transfer onto the compact object, and may trigger powerful episodes of jet activity. This scenario was used to explain the behaviour of a transient neutron star ULX (coincidentally also located in NGC\,300) associated with the SN impostor 2010da \citep[e.g.,\,][]{2016MNRAS.457.1636B,2016ApJ...830...11V,2018MNRAS.476L..45C}. One of the hallmarks of such a scenario is that the erupting donor star would be surrounded by thick circumstellar dust, and would appear highly reddened \citep[e.g.,\,][]{2009ApJ...699.1850B,2016ApJ...830..142L}. We mentioned earlier (Section \ref{sc:stellar_count}) that one of the candidate optical counterparts, star D, is indeed highly reddened. We leave a detailed investigation of this possibility to further work. 
\newline\\
\noindent {\it (v) Sheath-spine jet:}\\
A stratified jet has previously been used to explain the emission from gamma-ray bursts (GRBs) \citep{2001ApJ...556L..37M,2002MNRAS.337.1349R,2003ApJ...594L..23V}. Known as a `spine-sheath' jet model, a fast moving core (spine) is surrounded by a slower moving outflow (sheath) (see Fig 1. in \citealt{2013ApJ...777...62I}). This is thought to result from the GRB jet punching through the progenitor envelope, entraining the material to create the sheath \citep{2001ApJ...556L..37M,2002MNRAS.337.1349R}, or from a decoupled neutron sheath and proton jet core \citep{2003ApJ...594L..23V}. For S\,10, we do not expect direct emission from the jet, but instead suggest that a similar stratified jet may be causing multiple hotspots via external shocks. As the jet propagates out, the slower moving outer layer (or layers) shock the ISM, creating the inner knots, while the faster moving core punches through, decelerating further downstream, creating the outer knots. 
\newline\\
We cannot discriminate between interpretations iii--v based on our current observations. Differentiating between a persistent or transient jet is important because it will help us identify the true power of S\,10. In the following section we attempt to estimate the time averaged jet power. However, if the jet does switch on and off, then our jet power will be a lower limit; because we do not know the duty cycle of S\,10, we cannot determine the instantaneous power of the jet. 

\subsection{Jet power}

While S\,10 has a unique morphology among Galactic and extragalactic stellar-mass objects, we relate it most closely to super-Eddington microquasars and ULXs such as SS\,433, S\,26 and Ho II X-1. These sources all demonstrate the presence of powerful jets either through radio/X-ray knots, shocked emission from jet/ISM interactions and/or extended radio nebulae. S\,10 has a 5.5\,GHz radio luminosity $L_{\rm 5.5\,GHz}\approx1\times10^{34}\,\ergs$, which is a factor of two greater than the radio luminosity seen in the SS\,433/W\,50 complex \citep{1986MNRAS.218..393D,1998AJ....116.1842D}. On the other hand, it is an order of magnitude less radio luminous than the microquasars NGC\,7793 S\,26 \citep{2010MNRAS.409..541S}) and M\,83 MQ1 \citep{2014Sci...343.1330S}. The physical size of the radio nebula of S\,10 ($\approx170\times55\,\parsec$) is also comparable to the bubbles of S\,26 ($\sim300\times150\,\parsec$; \citealt{2010MNRAS.409..541S}), Ho II X-1 ($\sim81\times40\,\parsec$; \citealt{2012ApJ...749...17C}) and SS\,433 ($\sim100\times50\,\parsec$; \citealt{1998AJ....116.1842D}). Thus, in this section, we primarily compare S\,10 to these super-Eddington sources and their jet-inflated radio nebulae.

Assuming equipartion between the magnetic field and energy in relativistic particles, we can determine the minimum energy conditions for synchrotron radiation. The minimum energy $W_{\rm min}$ can be expressed as
\begin{equation}
    W_{\rm min} \approx 3.0\times10^{13}\eta^{4/7}V^{3/7}\nu^{2/7}L_{\nu}^{4/7}\,\erg, \label{min_syn}
\end{equation}
\citep{2011hea..book.....L} where $\eta-1$ is the ratio of energy in baryons to that in relativistic electrons, $V$ is the bubble volume in m$^3$, $\nu$ is the observing frequency in Hz and $L_{\nu}$ is the luminosity density in $\watt\,\hertz^{-1}$. The elongated structure of S\,10 has dimensions of $\approx170\times55\,\parsec$. If we assume a cylindrical shape, this equates to a total volume of $\approx10^{55}\,\meter^3$. Using this, and our 5.5\,GHz luminosity density of $L_{\nu}=2.2\times10^{17}\,\watt\,\hertz^{-1}$, we calculate a synchrotron minimum-energy condition $W_{\rm min}=5.5\E{49}\eta^{4/7}\,\erg$ corresponding to a minimum magnetic field strength of $B_{\rm min}=7\,\mu\guass$. This is comparable to the minimum energies required to power the extended radio nebulae in other super-Eddington microquasars, such as NGC\,5408 X-1 ($\sim$10$^{49}$ erg: \citealt{2007ApJ...666...79L}; or $\sim$10$^{50}$ erg: \citealt{2006MNRAS.368.1527S}, depending on different assumptions for the electron energy distribution), IC 342 X-1 ($\approx9.2\E{50}\erg$: \citealt{2012ApJ...749...17C}), and Ho II X-1 ($\approx2.6\E{49}\erg$: \citealt{2012ApJ...749...17C}).\footnote{Assumes $\eta=1$.} However, we note that there are several unknowns associated with our analysis. Firstly, without the viewing angle, we can only estimate the volume of the radio bubble based on its shape projected on the sky. The filling factor of the magnetic field throughout the bubble is also unknown. Additionally, the uncertainty on the minimum and maximum energy of the electron population, and therefore on the spectral index above and below the ATCA observing bands, also contribute to the uncertainty of the minimum-energy estimate.

We can estimate the jet power using the method outlined by \cite{2010Natur.466..209P}. For a shock-ionized gas bubble, we know from standard theory \citep{1977ApJ...218..377W} that the total radiative luminosity $L_{\rm rad}$ is $\approx$27/77 of the mechanical power that is inflating the bubble. The total radiative luminosity of the cooling plasma can be determined from the luminosity in suitable diagnostic lines; for example, the H$\beta$ line emission (derived from Equations 2.2 and 2.4 of \citealt{1995ApJ...455..468D}, or from the shock-ionization code {\sc mappings v}: \citealt{2008ApJS..178...20A}) is a good proxy for the total luminosity, because its relative contribution does not depend too strongly on the shock velocity, or the metallicity of the shocked gas, or the density of the undisturbed ISM. For a plausible range of shock velocities $\approx$150--300 km s$^{-1}$, the total H$\beta$ luminosity of shocked gas plus precursor is $\approx$(2--3)$\times 10^{-3}$ times the input mechanical power. For an H$\beta$ luminosity $L_{H\beta} \approx (6\pm1) \times 10^{36}$ erg s$^{-1}$, this corresponds to a long-term-average jet power $P_{\rm kin} \approx$ (2--3) $\times 10^{39}$ erg s$^{-1}$, an order of magnitude lower than the values estimated for S\,26 but the same order of magnitude as the jet power suggested for SS\,433 \citep{2000A&A...363..640B,2002ApJ...564..941M,2009MNRAS.394.1674K} and for M\,51 ULX-1 \citep{2018MNRAS.475.3561U}. In a follow-up paper currently in preparation, we will refine and improve this estimate of jet power based on diagnostic line emission, using recently obtained data from the Multi Unit Spectroscopic Explorer (MUSE), an integral field spectrograph mounted on the VLT.



Other techniques have sometimes been used to estimate the kinetic power in radio/optical microquasar bubbles \citep{2010Natur.466..209P}. If the shock velocity is known (inferred from optical spectroscopic studies of the emission line profiles), together with the ISM density and the bubble size, one can derive the characteristic dynamical age of the shocked bubble as well as the jet power \citep{1977ApJ...218..377W}. A spectroscopic study of the optical emission lines in S\,10 is also left to our follow-up paper based on MUSE data. A much less reliable technique is based on the radio luminosity. Various scalings between the jet power and the optically thin synchrotron luminosity of hot spots, lobes and cavities have been proposed, for samples of radio galaxies \citep[e.g.,\,][]{1999MNRAS.309.1017W,2010ApJ...720.1066C,2013ApJ...767...12G}. 
The model assumptions underpinning these scaling relations are similar to those associated with the minimum energy synchrotron conditions discussed earlier, leading to a similar degree of uncertainty. However, \cite{2016MNRAS.456.1172G} have shown that the scaling relations commonly used in the literature are strongly biased by the mutual dependence on distance for both the jet power and the radio luminosity. Thus, we do not rely on those correlations for our study.

\subsection{Synchrotron or thermal plasma emission?} \label{sc:synch_vs_thermal}

From the synchrotron minimum energy conditions, we can estimate the cooling timescale for X-ray synchrotron emission. At an energy of 1\,\kev, the cooling timescale would be $3,000\left(\frac{B_{\rm min}}{7\,\mu\guass}\right)^{-3/2}\,\yr$ \citep{2006MNRAS.372..417T}. This approximation is valid if the system is in equipartition; a stronger magnetic field will reduce the cooling timescale. Adiabatic expansion of the knots would also lead to faster cooling.

Based on the $5.5\,\GHz$ luminosity and average spectral index ($\alpha=-0.64$) for the radio bubble, the expected $0.3-8.0\,\kev$ X-ray synchrotron luminosity is $\approx3\E{37}\,\ergs$. We have already tested (Section \ref{xray_knots_results_sec}) that a power-law component with a photon index $\Gamma=1.7$ does not provide a statistically significant contribution to the combined knot spectrum, and the 90\% upper limit to its luminosity is $<2\E{36}\,\ergs$. For completeness, given the spectral index found in the radio analysis (Section \ref{sc:radio_results}) and that we now suspect the microquasar candidate core resides within X-ray knot 2 (Section \ref{sc:xray_core}), we perform the test with an X-ray power-law component of photon index $\Gamma=1.64$ on the individual spectrum of knot 2. As expected, the result is essentially identical to the case for the combined spectra and for $\Gamma=1.7$, that is there is no statistically significant need for such a component, and the 90\% upper limit to its contribution is $<8\E{35}\,\ergs$. This is lower than the luminosity expected from the extrapolation of the radio synchrotron power law.
The reason why we do not see synchrotron X-ray emission may be that particles were never accelerated up to energies where they could emit X-rays via this process; or there may be a break in the spectrum between radio and X-ray frequencies because of ageing. We also note that the radio and X-ray observations were not simultaneous. Alternatively, the S\,10 core may be producing X-ray synchrotron emission that is strongly absorbed to the point where it is not detected. While the thermal X-ray emission does not show any evidence of strong absorption, it is likely a result of extended jet/ISM interactions, whereas the X-ray synchrotron emission must emanate from a region much closer to the compact object and could be obscured by an additional strong source of absorption; $n_{\rm H}>4\E{23}$ atoms cm$^{-2}$. This scenario seems somewhat contrived and thus we prefer the simpler interpretation wherein the synchrotron power-law has a break below the soft X-ray band, in line with other microquasars \citep{2013MNRAS.429..815R}, and both the radio synchrotron and the X-ray thermal emission come from similar regions of jet/ISM interactions.



While we do not detect any X-ray synchrotron emission, we do see thermal X-ray emission (Bremsstrahlung and line transitions), likely from shock-heated gas along and in front of the jet. The cooling timescale is approximately the ratio of total heat content over heat loss rate 
\begin{equation}
    t_{\rm cool} = \frac{3(n_{\rm e}+n)k_{\rm B}T}{2n^2\Lambda},\label{eq:cool}
\end{equation}
\citep{2003adu..book.....D} where $n$ is the atomic density of the hot gas, $k_{\rm B}$ is the Boltzmann constant, $T$ is the plasma temperature and $\Lambda$ is the cooling function. From our X-ray spectral fits, we know that $T\approx 0.6\,\kev\approx 7 \times 10^6\,\Kelvin$. At that temperature, $\Lambda \approx 4 \times 10^{-23}$ erg cm$^3$ s$^{-1}$ for solar-metallicity plasma in collisional equilibrium, including both free-free and line emission \citep{1993ApJS...88..253S}. From the thermal X-ray luminosity and the approximate volume estimated for the emitting gas, we also determined a lower limit on the density, $n \approx n_e \gtrsim 4\,\centi\meter^{-3}$ (Section \ref{xray_knots_results_sec}). Putting those constraints together, we obtain an upper limit on the (thermal) cooling timescale of $t_{\rm cool} \lesssim 570,000\,$yr. We speculate that X-ray-emitting gas is replenished by bursts of increased accretion/ejection activity leading to the production of new discrete knots on a recurrence timescale shorter than the cooling timescale.



\subsection{Faintness of the X-ray core}

Three possibly super-Eddington microquasars have been found so far with large-scale X-ray evidence of collimated jets: NGC\,300 S\,10, NGC\,7793 S\,26 and SS\,433 in the Milky Way. Curiously, all three have apparently faint cores: $L_{\rm {X,core}} \lesssim 10^{36}$ erg s$^{-1}$ (Section 3.1.2), $L_{\rm {X,core}} \approx6\E{36}\,\ergs$ \citep{2010MNRAS.409..541S} and $\sim10^{36}\,\ergs$ \citep{1996PASJ...48..619K,2011IAUS..275..280F} for the three sources, respectively. In the case of SS\,433, the reason is the occultation of the direct X-ray emission by the thick super-critical disk, seen at high inclination \citep{2004ASPRv..12....1F}. For S\,26 and S\,10, we do not have any estimate of the viewing angle. No firm conclusions can be reached based on such a small sample of objects. It is possible that there is a selection bias at play: large-scale jets appear longer and easier to discover when they are in the plane of the sky; as a result, the accretion disk around the compact object would preferentially appear edge-on. The situation would be analogous to that of ``accretion disk corona'' sources, a type of Galactic low-mass X-ray binary seen at high inclination, in which the direct X-ray emission from the disk surface is mostly occulted from us, and we can only see harder radiation scattered by a vertically extended corona \citep[e.g.,\,][]{1982ApJ...257..318W,1982ApJ...258..245M,1989MNRAS.239..715H}. Alternatively, the compact object may be faint because its super-critical accretion phase has a short duration compared with the cooling timescale of the hot spots and bubble, or it is a transient source with a low duty cycle. Contrary to this scenario, it was noted \citep{2008AIPC.1010..303P} that the majority of large shock-ionized ULX bubbles (without direct evidence of a collimated jet) do contain a luminous X-ray core; if the typical duty cycle of super-Eddington accretors was low, or if the duration of the super-Eddington phase was much shorter than the cooling timescale of the ULX bubbles, we would see a large number of bubbles without a central source, which is not the case \footnote{A similar argument also rules out beaming factors larger than a few for the X-ray emission of ULXs \citep{2008AIPC.1010..303P}.}.

We can take the low-duty-cycle argument to an extreme, and suggest that all the kinetic power in the collimated jet was injected at an instantaneous burst event, for example a SN associated with a Gamma-Ray-Burst (GRB). After that event, the collimated ejecta would continue to expand passively, now dominated by entrained material from the ISM. This scenario leads us back to the original interpretation of S\,10 as an SNR. Magneto-hydrodynamical simulations of jet/ISM interactions in GRBs suggest \citep{2012ApJ...751...57D} that the jet loses collimation at a distance of $\sim$1 pc from the origin; this makes it difficult to explain the much larger length of the S\,10 structure. To test this scenario further, we searched for evidence of current accretion activity at the (candidate) location of the central object, as explained in the next section.

\subsection{Search for X-ray and radio variability} \label{var_sect}

Spurred by the possibility that the central engine of S\,10 is a transient or variable accretor, currently in a low state, we searched for hints of previous variability in the core emission. 
S\,10 has been observed on several occasions with the {\it ROSAT} and the {\it XMM Newton} X-ray telescopes: in both cases, the source was not resolved, because of their poorer spatial resolution (see, {\it e.g.}, Figures 8e and 8f in \citealt{2004A&A...425..443P}). 
For the {\it ROSAT} observations, we used the count rates listed in the Second  Position-Sensitive Proportional Counter (PSPC) Catalog \citep{2000yCat.9030....0R} and in the Source Catalog of Pointed Observations with the High Resolution Imager (HRI) \citep{2000yCat.9028....0R}. For the {\it XMM-Newton} observations, we used the observed fluxes from the 3XMM-Data Release 6 Catalog \citep{2016yCat.9047....0R}. 
We then used the online tool {\sc PIMMS}\footnote{http://cxc.harvard.edu/toolkit/pimms.jsp} version 4.8e to convert the count rates or fluxes of the various observations to unabsorbed 0.3--8 keV fluxes and then to luminosities, assuming a $0.6\,\kev$ thermal plasma model and a column density of $n_{\rm H}=3\E{20}\,\centi\meter^3$, as for our spectral modelling (Section \ref{xray_results_sec}, Table \ref{xray_tab}). 

For the {\it ROSAT}/PSPC observations between 1991 November and 1992 January, we estimate a luminosity $L_{0.3-8} = (1.3 \pm 0.3) \times 10^{37}$ erg s$^{-1}$; for the PSPC observations between 1992 May--June, $L_{0.3-8} = (1.1 \pm 0.2) \times 10^{37}$ erg s$^{-1}$.
For the {\it ROSAT}/HRI observation in 1995 May, we obtain a luminosity $L_{0.3-8} = (1.5 \pm 0.4) \times 10^{37}$ erg s$^{-1}$. Finally, {\it XMM-Newton} observed NGC\,300 between 2000 December and 2001 January; the absorbed 0.2--12 keV flux of $(3.44 \pm 0.13) \times 10^{-14}$ erg cm$^{-2}$ s$^{-1}$ converts to an emitted 0.3--8 keV luminosity $L_{0.3-8} = (1.6 \pm 0.2) \times 10^{37}$ erg s$^{-1}$. Considering the model uncertainties and the differences between the PSFs and energy bands of the various detectors, we cannot claim any significant X-ray variability compared with our modelled {\it Chandra} luminosity $L_{0.3-8} = (1.1 \pm 0.1) \times 10^{37}$ erg s$^{-1}$ (average of the 2010 and 2014 observations). Any possible variability of the core was swamped by the higher and constant thermal emission from the various knots. 

\begin{table}
    \centering
    \caption{Integrated radio flux of the core region at various epochs.} 
    \footnotesize
    \begin{tabular*}{0.48\textwidth}{l @{\extracolsep{\fill}} rrrrr}
        \hline\hline
        \multicolumn{1}{c}{Telescope} & \multicolumn{1}{c}{Year}&  \multicolumn{1}{c}{$\nu$} & \multicolumn{1}{c}{Flux density} & \multicolumn{1}{c}{$\nu L_{\nu}$}\\
         & &  \multicolumn{1}{c}{($\GHz$)} &  \multicolumn{1}{c}{($\milli\jansky$)} &  \multicolumn{1}{c}{($10^{33}\,\ergs$)}\\
        \hline
        ATCA & 2000-02-28 & 1.374 & $0.56\pm0.03$ & 3.3 \\
        ATCA & 2000-02-28 &  2.496 & $0.38\pm0.06$ & 4.0 \\
        ATCA$^*$ & 2015/16 & 5.5 & $0.429\pm 0.030$ & 10.0 \\
        ATCA$^*$ & 2015/16 &  9.0 & $0.352\pm0.029$ & 13.4\\        
        \hline
    \end{tabular*}
    \begin{flushleft}
    $^*$ ATCA observations from this study (October 2015 and August 2016) are stacked as per Section \ref{rad_dat_red_sec}.
    \end{flushleft}
    \label{rad_var_tab}
\end{table}

We then checked whether S\,10 has varied in radio brightness over the last decades. NGC\,300 was observed with the VLA at $4.885\,\GHz$ in May 1993 and at $1.465\,\GHz$ in June 1998 \citep{2000ApJ...544..780P}. It was also observed with the ATCA in February 2000 at $1.374\,\GHz$ and $2.496\,\GHz$ \citep{2004A&A...425..443P}. We compared these data with our recent 2016--2017 ATCA observations. One major difficulty in this comparison is that the archival data have larger beam sizes: $6'' \times 6''$ for the old ATCA observations, $4''.70 \times 3''.76$ for the 1.465\,GHz VLA observations, and $8''.63 \times 4''.22$ for the 4.885\,GHz VLA observations (Table 2 in \citealt{2004A&A...425..443P}). This makes it difficult to distinguish between the radio flux from the S\,10 jet and that from the neighbouring H\,10 H{\sc ii} region. The second major problem is that the archival observations were much less sensitive. The rms noise level was $\sigma \approx 37 \mu$Jy beam$^{-1}$ and $\sigma \approx 66 \mu$Jy beam$^{-1}$ for the VLA data at $4.885\,\GHz$ and at $1.465\,\GHz$, respectively; for the archival ATCA data it was $\sigma \approx 58 \mu$Jy beam$^{-1}$ and $\sigma \approx 62 \mu$Jy beam$^{-1}$ at $1.374\,\GHz$ and $2.496\,\GHz$, respectively. As a comparison, the rms noise was $\sigma \approx 2.7 \mu$Jy beam$^{-1}$ in the 2015--2017 ATCA data (Section 2.2). As a result, the emission from the core region (around the X-ray knot 2) is the only part of the radio jet that is significantly detected in the archival data (see for example Fig.~8d of \citealt{2004A&A...425..443P}). Keeping in mind all these caveats, and extrapolating the 2015--2016 flux measurements to lower frequencies using a spectral index $\alpha = -0.40\pm0.22$ for the core region, we find that the new ATCA flux measurements are consistent with the 2000 ATCA flux values within $\approx$2$\sigma$. Instead, when compared with the VLA flux densities measured by \cite{2000ApJ...544..780P} and \cite{2004A&A...425..443P}, S\,10 appears to have brightened by a factor of 2 in the ATCA observations; however, our re-analysis of the VLA data shows that the true uncertainties in the S\,10 core flux are much larger than the formal error reported by \cite{2000ApJ...544..780P} and \cite{2004A&A...425..443P}, because of the mismatch in beam size and sensitivity mentioned above. To combat this, and properly compare the S\,10 core fluxes, we re-analysed both the 4.8\,GHz VLA data and our 5.5\,GHz ATCA data. These observations were chosen to minimise uncertainties introduced by the spectral shape of S\,10. Additionally, we applied the same \textit{uv}-cut to both data sets and used the same restoring beam. We find peak fluxes of $f_{4.8\,\GHz}=340\pm60\,\ujy\,$\perbeam and $f_{5.5\,\GHz}=210\pm8\,\ujy\,$\perbeam. We find tentative evidence ($\sim2\sigma$) of a decrease in the flux of the radio core, though future observations are required to conclusively establish the presence of variability.

\section{Conclusions}

We have presented a new, coherent interpretation of a complex X-ray, optical and radio source in NGC\,300, using new and archival {\it Chandra}, {\it HST}, VLT and ATCA data. The source was previously classified as an SNR; however, we argued that it is a candidate microquasar, likely powered by super-Eddington accretion onto a compact object. We showed that the X-ray emission is made of a string of discrete knots, a tell-tale sign of a jet interacting with the ambient medium. We also showed for the first time that the radio emission is an elongated bubble ($\approx$170 $\times 55$ pc in size), oriented along the same direction as the string of X-ray knots. While the radio emission is consistent with optically-thin synchrotron (as expected), the X-ray emission is from thermal plasma at temperatures $\approx$0.3--0.8 keV. This is unlike the knots and hot spots typically seen in AGN jets, but is analogous to the thermal hot spots previously identified in the super-Eddington microquasar NGC\,7793 S\,26. The integrated radio luminosity at 5.5 GHz is $\approx$10$^{34}$ erg s$^{-1}$, while the total X-ray luminosity of the knots is $\approx$2 $\times 10^{37}$ erg s$^{-1}$. To complete the picture, we argued that the optical line spectrum of the shock-ionized nebula is exactly what is expected for this kind of system, as seen also in S\,26 and in other shock-ionized ULX bubbles. In particular, we showed a spatial association between locations of enhanced H$\alpha$ emission and locations of X-ray and radio features.

Using the H$\alpha$ line luminosity ($L_{H\alpha} \approx 1.7 \times 10^{37}$ erg s$^{-1}$) as a proxy, we estimated that the nebula was shocked by the injection of a long-term-average jet power $P_{\rm {kin}} \approx$ (2--3) $\times 10^{39}$ erg s$^{-1}$. Thus, this candidate microquasar is in the same energy class as SS\,433 and Ho II X-1; it is an order of magnitude less energetic and a factor of 2 smaller in linear size than S\,26. 

We have discussed the most likely location of the core, but found no bright, point-like X-ray source there, above the thermal-plasma emission. The upper limit to the X-ray luminosity of the core is $\approx$10$^{36}$ erg s$^{-1}$. The direct emission from the accreting compact object may be occulted from our view by a thick, edge-on disk, like in SS\,433; or the core could be in a low state. Neither did we find evidence of a brighter core in earlier {\it XMM-Newton} and {\it ROSAT} observations, which stretch back to 1991. The reason for the multiple knot structure remains unclear. We can rule out internal shocks  (analogous to the knots seen for example in the M\,87 jet), because the X-ray emission is thermal (that is, from shocked ISM) rather than synchrotron. For the same reason, we consider very unlikely that the knots are ballistic ejections. We discussed alternative scenarios, in particular: multiple layers of enhanced ISM density; multiple outbursts or jet activity episodes; or a sheath-spine jet structure, with the faster spine propagating a longer distance than the slower sheath. 

In summary, S\,10 is a key object to understand jets in the super-critical accretion regime, and is an ideal source for follow-up multiwavelength observations. The most important piece of information that we are missing is the shock velocity of the optically-emitting gas, including possible velocity differences between the plasma above and below the core position. This would provide us with a dynamical age of the bubble and an alternative measurement for the mechanical power. We will tackle this problem in follow-up work currently in preparation, based on recently obtained MUSE data; thanks to those new results, we will be able to directly compare the estimates of jet power inferred from the kinematics of the shocked gas, and from the optical line fluxes. 
Increasing the sample size of super-Eddington microquasars (and, more generally, of shock-ionized ULX bubbles) is crucial for our understanding of jet launching and collimation processes, and energy feedback. 

\section*{Acknowledgements}

We thank William Blair, Jifeng Liu, Vlad Tudor and Sam McSweeney for useful discussions. We also thank the anonymous referee whose constructive feedback helped to improve this paper. RU acknowledges that this research is supported by an Australian Government Research Training Program (RTP) Scholarship. RS acknowledges support from a Curtin University Senior Research Fellowship; he is also grateful for support, discussions and hospitality at the Strasbourg Observatory during part of this work. JCAM-J is the recipient of an Australian Research Council Future Fellowship (FT140101082). GEA is the recipient of an Australian Research Council Discovery Early Career Researcher Award (project number DE180100346) funded by the Australian Government. RMP acknowledges support from Curtin University through the Peter Curran Memorial Fellowship. The International Centre for Radio Astronomy Research is a joint venture between Curtin University and the University of Western Australia, funded by the state government of Western Australia and the joint venture partners. The scientific results reported in this article are based data obtained from the Chandra Data Archive. This research has made use of software provided by the Chandra X-ray Center (CXC) in the application package CIAO. Based on observations made with the NASA/ESA Hubble Space Telescope, obtained from the Data Archive at the Space Telescope Science Institute, which is operated by the Association of Universities for Research in Astronomy, Inc., under NASA contract NAS 5-26555. These observations are associated with program 10915. IRAF is distributed by the National Optical Astronomy Observatories, which are operated by the Association of Universities for Research in Astronomy, Inc., under cooperative agreement with the National Science Foundation. This research has made use of the VizieR catalogue access tool, CDS, Strasbourg, France. This work has made use of data from the European Space Agency (ESA) mission {\it Gaia} (\url{https://www.cosmos.esa.int/gaia}), processed by the {\it Gaia} Data Processing and Analysis Consortium (DPAC, https://www.cosmos.esa.int/web/gaia/dpac/consortium). Funding for the DPAC has been provided by national institutions, in particular the institutions participating in the {\it Gaia} Multilateral Agreement. The Australia Telescope Compact Array is part of the Australia Telescope National Facility which is funded by the Australian Government for operation as a National Facility managed by CSIRO. This research has made use of NASA's Astrophysics Data System.

\bibliography{references}

\bsp	
\label{lastpage}
\end{document}